\documentclass[%
reprint,
superscriptaddress,
amsmath,amssymb,amsfonts,
aps,
prb,
floatfix,notitlepage
]{revtex4-1}

\usepackage[dvipdfmx]{graphicx}
\usepackage{dcolumn}
\usepackage{bm}
\usepackage{braket}
\usepackage{mediabb} 
\usepackage[dvipdfmx]{attachfile2}
\usepackage{ulem}

\begin{document}

\title{Accessing electromagnetic properties of matter with cylindrical vector beams}

\author{Hiroyuki Fujita}
\email{h-fujita@issp.u-tokyo.ac.jp}
\thanks{Corresponding author}
\affiliation{Institute for Solid State Physics, University of Tokyo, Kashiwa 277-8581, Japan}
\author{Yasuhiro Tada}
\email{tada@issp.u-tokyo.ac.jp}
\affiliation{Institute for Solid State Physics, University of Tokyo, Kashiwa 277-8581, Japan}
\author{Masahiro Sato}
\email{masahiro.sato.phys@vc.ibaraki.ac.jp}
\affiliation{Department of Physics, Ibaraki University, 
Mito, Ibaraki 310-8512, Japan}
\date{\today}

\begin{abstract}
Cylindrical vector beam (CVB) is a structured lightwave characterized by its topologically nontrivial nature of the optical polarization. The unique electromagnetic field configuration of CVBs has been exploited to optical tweezers, laser accelerations, and so on. However, use of CVBs in research fields outside optics such as condensed matter physics has not progressed. In this paper, we propose potential applications of CVBs to those fields based on a general argument on their absorption by matter. We show that pulse azimuthal CVBs around terahertz (THz) or far-infrared frequencies can be a unique and powerful mean for time-resolved spectroscopy of magnetic properties of matter and claim that an azimuthal electric field of a pulse CVB would be a novel way of studying and controlling edge currents in topological materials. We also demonstrate how powerful CVBs will be as a tool for Floquet engineering of nonequilibrium states of matter. 
\end{abstract}

\maketitle

\section{Introduction}

Developments of optics in the past decades enabled the realization of high-intensity, ultra-short pulse lasers~\cite{STRICKLAND1985219,doi:10.1063/1.1148795}. The high temporal resolution due to the short pulse width allows us to directly access to nonequilibrium phenomena like chemical reactions and photo-induced phenomena in matter. The high intensity opened new fields of photo-induced physics~\cite{doi:10.1142/5476}, ultrafast magnetism~\cite{RevModPhys.82.2731}, and femtochemistry~\cite{Zewail1645}. 

Moreover, the intense ultra-short light becomes vital for generating intense terahertz (THz) pulses~\cite{Tonouchi:2007aa}. Since THz is the typical energy scale of various collective phenomena~\cite{George:2008aa,Liu:2012aa,PhysRevLett.111.057002,Matsunaga1145} in solids like superconductivity, phonons, and magnons and is also crucial for understanding biomolecules~\cite{Markelz:2000aa,0031-9155-47-21-319,0031-9155-47-21-318,F.:2007aa}, the developing THz optics shares a central position in optical physics.

Existing studies of optical phenomena in matter have been mainly done with simple Gaussian beams, and  much attention has not been paid to other kinds of lasers developed in the past decades in condensed matter physics and chemistry. Among those modern variants of lasers, here we focus on so-called structured lights. In particular, we discuss applications of cylindrical vector beams (CVBs)~\cite{Youngworth:00,Zhan:02,Zhan:09}. 

Because of their unique focusing property, CVBs potentially have substantial impacts on studies of optical phenomena in matter. Nevertheless, unlike other structured lights like optical vortices~\cite{PhysRevA.45.8185,9780511795213}, applications of CVBs outside optics have not progressed.  In particular, the magnetic field component of CVBs, which we mainly discuss in this paper, has been overlooked. There are important studies using CVBs for the spectroscopy of molecules and quantum dots~\cite{Lieb:01,Lieb:04,PhysRevLett.96.113002,Gutbrod:2010aa,Zuchner:2011aa} and for the excitation of plasmonic modes of metal nanostructures~\cite{Sancho-Parramon:2012aa,Wackenhut:2012aa}. However, all those experiments are about the electric field component of CVBs. 
They have been indeed fruitful, but it is desired to explore applications of the magnetic field component.

In this paper, we show that using CVBs instead of conventional Gaussian beams can be a quite powerful way of studying magnetic properties of matter especially in the THz or far-infrared frequency ranges. By using measurement samples sufficiently smaller than the wavelength, which can be easily achieved nowadays, we can control the applied electric and magnetic fields almost independently. In particular, we can avoid strong electric dipole absorption which often overwhelms magnetic absorption which we are interested~\cite{PhysRevX.7.011005}.

Using THz CVBs would be particularly effective for time-resolved measurements. The relatively short timescale of THz CVBs offers high temporal resolution, and the characteristic spatial profile of focused CVBs allows us to probe electric and magnetic properties of matter selectively. We will give several applications of CVBs utilizing these properties; time-resolved electron spin resonance (ESR) of conducting materials, a novel way of studying multiferroic materials, electron paramagnetic resonance (EPR) study of the dynamics of biomolecules in their living environments. In addition to them, we consider how to exploit the characteristic azimuthal electric field configuration of CVBs for probing and controlling edge currents in topological materials~\cite{RevModPhys.82.3045,RevModPhys.83.1057}. 

We also discuss applications of focused CVBs for Floquet engineering~\cite{doi:10.1080/00018732.2015.1055918,RevModPhys.89.011004,kitamura-oka}. We demonstrate that the highly controllable nature of the electromagnetic fields of CVBs offers a building block for realizing desired nonequilibrium physics in periodically driven matter. We take a simple model of quantum magnets and compare the Floquet engineering with CVBs and that with circularly polarized lasers~\cite{arXiv:1602.03702,higashikawa}.

The rest of this paper is organized as follows. In the second section, we review CVBs. In the third section, we study how small the electric field absorption can be by replacing Gaussian beams with CVBs. We give potential applications of CVBs for condensed matter physics and chemistry$/$biology.  In the forth section, we deal with the electric field component of CVBs and show that the electric field absorption for a pulse azimuthal CVB can be a way of probing and controlling edge states of topological matter. The fifth section is about Floquet engineering. We show that CVBs have clear advantages over conventional lasers for designing nonequilibrium states of matter. The final section is devoted to the conclusion.

\section{Cylindrical vector beam}\label{sec:CVB}
Exploring applications of topological lightwaves like optical vortices~\cite{PhysRevA.45.8185,9780511795213} and CVBs~\cite{Youngworth:00,Zhan:02,Zhan:09} is one of the central issues in modern optics. Both beams have vanishing intensity along the propagation axis with different physical origins. In the case of optical vortices, their scotoma comes from the spiral-shaped phase structure. The azimuthally inhomogeneous wavefront carries a non-vanishing orbital angular momentum (not a spin angular momentum). On the other hand, CVBs are azimuthally homogeneous. The topological nature of CVBs originates from the spatial profile of the polarization vector. That is, the scotoma is nothing but the vortex core of the in-plane components of the polarization vector of CVBs. Both beams are experimentally realized in the broad range of frequencies by using holograms or structured filters~\cite{9780511795213,Zhan:09}. 

There are a number of applications of optical vortices in optics, condensed matter physics, biology, and so on~\cite{Rittweger:2009aa,Hamazaki:10,Terhalle:11,Toyoda:2012aa,Takahashi:13,PhysRevB.93.045205,Fujita2016,PhysRevB.96.060407} including the stimulated emission depletion microscopy~\cite{Rittweger:2009aa} which is awarded the Novel prize in chemistry in 2014. Compared to optical vortices, however, the use of CVBs outside optics is, not well explored~\footnote{Very recently, two of the authors proposed its application for the Fermi surface measurement of magnetic metals and generic metals under ultra-high pressure~\cite{Fujita:2018aa}}.

\begin{figure}[htbp]
\centering
\includegraphics[width = 80mm]{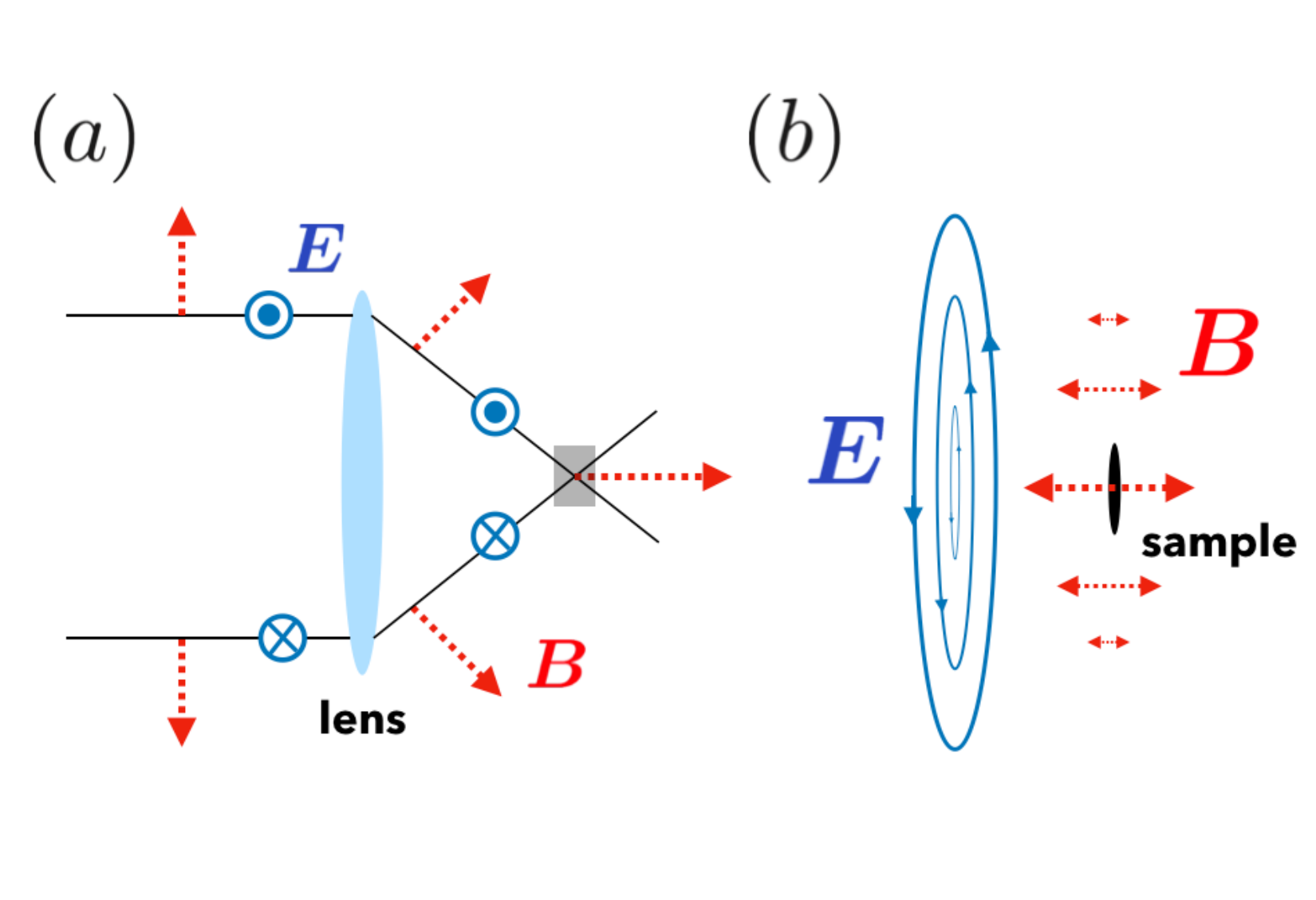}
\caption{$(a)$: Schematics of the tightly focused azimuthal CVB. Arrows indicate the direction of electromagnetic fields. $(b)$: Close view of the shaded region in $(a)$. In this paper, we consider the case where the sample size is much smaller than the wavelength. Near the focus, the azimuthal component of the beam is small while the longitudinal component of the magnetic field becomes prominent. }
       \label{setup}
  \end{figure}

Let us give a mathematical description of CVBs following a literature~\cite{Youngworth:00}. A CVB is obtained by solving the Maxwell's equation in a vacuum in the cylindrical coordinate $(\rho,\phi,s)$. Here, $\rho$ is the radial coordinate, $\phi$ is the azimuthal angle around the cylindrical axis, and $s$ is the coordinate along the cylindrical axis. On the focal plane, an azimuthal CVB (in the following, we mostly omit ``azimuthal'') with the wave number $k$ and the frequency $\omega = ck$ is given by:

\begin{align}\label{CVBeq}
E^\phi(\rho,\phi) &= 2 A \int^{\alpha}_{0}  \sin\theta \cos^{\frac{1}{2}}(\theta)\ell_0 (\theta) J_1(k\rho \sin\theta)  d\theta, \nonumber \\
B^\rho(\rho,\phi) &= -\frac{2}{i c} A \int^{\alpha}_{0}  \sin\theta\cos^{\frac{3}{2}}(\theta)\ell_0 (\theta) J_1(k\rho \sin\theta)  d\theta, \nonumber\\
B^z(\rho,\phi) &= \frac{2}{i c} A \int^{\alpha}_{0} \sin^2\theta \cos^{\frac{1}{2}}(\theta)\ell_0 (\theta) J_0(k\rho \sin\theta)  d\theta 
\end{align}
where $c$ is the speed of light in the vacuum, $J_n(x)$ is the Bessel function, and $l_0(\theta)$ is the pupil apodization function determined by the optical system~\cite{Youngworth:00}. Other components of electromagnetic fields are zero. The size of the pupil $\alpha$ is given by $\alpha = \sin^{-1}({\rm NA}/n)$ where $n$ and NA are the refractive index and the numerical aperture of the lens, respectively. The constant $A$ determines the field amplitude.
In Fig.~\ref{setup}, we show the schematics of a tightly focused CVB and its field configuration near the focus.

Notable features of focused CVBs are (i) vanishing amplitude of electric fields at the center ($\rho = 0$) and (ii) non-vanishing ``longitudinal'' magnetic field there. As a result, near the focus, the longitudinal magnetic field becomes the dominant component of the CVB. Therefore, if we shine a CVB on a tiny sample placed at the focus of the beam [Fig.~\ref{setup}($b$)], the sample is virtually under an oscillating longitudinal magnetic field at an optical frequency while the electric field components are kept small~\footnote{The electromagnetic field configuration of focused CVBs is qualitatively the same as the TE011 cavity mode of microwave~\cite{Hertel:2005aa} which is already the standard of ESR at microwave frequencies. Focused CVBs are, therefore, providing an optical way of mimicking the electromagnetic configuration of TE011 mode.}. As the focusing of the beam gets tighter, the longitudinal field becomes more prominent and the relative amplitude of the electric field component decreases.

To make use of the longitudinal field of a focused CVB, we have to prepare a sample much smaller than the wavelength of the incident beam. For visible lights, this requirement is hardly satisfied in the condensed-matter context, but for (sub-) THz or far-infrared lights, it is realistic. Indeed, by using focused ion beam equipment, it is nowadays possible to prepare a sample with its size 100 nm to $\mu$m. Therefore, if we use the (sub-) THz CVB (wavelength of the order of 100 $\mu$m to mm) we could measure the magnetic field absorption while the electric field one is strongly suppressed. In the next section, we quantitatively discuss this point.

\section{Longitudinal magnetic field: magnetic field spectroscopy}
In this section, we discuss usefulness of the longitudinal magnetic field of focused CVBs for spectroscopic studies of magnetic properties of matter. We will see that CVBs in the THz region is particularly useful as their relatively long wavelength enables us to strongly suppress the electric field absorption while keeping the high temporal resolution of the measurements.

\subsection{Suppressed electric field absorption for CVB}\label{suppress}
Electromagnetic absorption of matter is characterized by the electrical conductivity and magnetic susceptibility. In this subsection, to demonstrate that using CVBs instead of Gaussian beams results in the strong suppression of the electric field absorption, we examine a high-frequency ESR measurement of electric conductors.

In electron spin resonance (ESR)~\cite{wertz2012electron,slichter1996principles}, the imaginary part of the magnetic susceptibility, which is relevant for the magnetic field absorption, has a sharp peak at the frequency of magnetic dipole transitions. In the simplest case where the spin-splitting is caused by the Zeeman effect of the external static magnetic field $H_0$, the resonance frequency $\omega_0$ is  written as $\gamma H_0$ using the gyromagnetic ratio $\gamma$. In ESR, from the peak positions and line widths of ESR absorption spectra, we obtain the resonance frequency, spin relaxation time, magnetic anisotropy, and so on~\cite{wertz2012electron,slichter1996principles}.

In electric conductors, in addition to the magnetic field absorption, we always encounter strong electric field absorption determined by the real part of the electric conductivity. In many cases, the electric field absorption is order of magnitude stronger than the magnetic field one. Hence, detecting magnetic contribution to the absorption spectrum is challenging in conducting systems. A possible approach is a subtraction of the electric field contribution for example by using the phenomenological Drude-law~\cite{Ashcroft} to single out the magnetic contributions.
However, the absorption spectra of real materials are quite complicated, and the actual absorption spectra deviate from the Drude-law (we call those deviations microstructures in the following), making it hard or impossible to separate those two contributions out.

The typical frequency of the oscillating field in ESR measurements is gigahertz (GHz). Recently, however, high-frequency ESR using (sub-) Tera Hz (THz) fields is becoming important~\cite{0953-8984-12-47-201,0034-4885-68-2-R05,grinberg2013very}. High-frequency ESR extends the applicability of ESR to matter with a sizable zero-field spin splitting~\cite{Kampfrath:2011aa,PhysRevB.87.094422} such as antiferromagnets and spin liquids~\cite{0034-4885-80-1-016502}.  Moreover, the use of high-frequency (for example THz) fields improves the temporal resolution of the time-dependent ESR for dynamical properties of free radicals in the target.

At high frequency, however, the absorption by conduction electrons mentioned above becomes more serious. The dominant electric field absorption combined with microstructures in the electric conductivity is an obstacle for measuring ESR signals. For example, when there is a strong spin-orbit-coupling (SOC), so-called the electron dipole spin resonance (EDSR) takes place~\cite{PhysRevB.78.195302,Nowack1430}. As a result, the high and broad peaks originating from the EDSR mechanism~\cite{PhysRevB.95.235115} contribute to the conductivity microstructures and hide ESR peaks. Even without SOC, microstructures reflecting materials details could be problematic. Indeed, there has been no report of THz ESR for conducting materials so far.

Combining the electric and magnetic field absorption (coming for example from ESR), as a whole, we can write the electromagnetic absorption of a matter as~\footnote{Here we omit possible cross-terms of $\bm{E}$ and $\bm{B}$. For example, in multiferroic materials, they could be important.}
\begin{align}\label{absorb}
\alpha = \sum_{i, j} \sigma'_{ij} (\omega) E_i E_j + \omega \chi_{ij}''(\omega) B_i B_j.
\end{align}
Here $\sigma'$ is the real part of the electrical conductivity and $\chi''$ is the imaginary part of the magnetic susceptibility. The subscripts $i, j$ are for the spatial coordinates. Let us assume that our sample has the radius of $R \ll \lambda$ ($\lambda$ is the wavelength of the beam) and is placed at the focus of the CVB. We expand $E^\phi(\rho,\phi)$ in Eq.~\eqref{CVBeq} in terms of $\rho k$ satisfying $\rho k < R/\lambda \ll 1$ and obtain
\begin{align}\label{leadingorder}
E^\phi(\rho,\phi) =  2 A k\rho  \int^{\alpha}_{0}  \sin\theta \cos^{\frac{1}{2}}(\theta)\ell_0 (\theta) \frac{\sin\theta}{2}d\theta + O(k^3\rho^3).
\end{align}
We see that the leading order term is $O(k\rho)$, and that of the absorption Eq.~\eqref{absorb} is then $O(k^2\rho^2)$. The overall absorption is obtained by integrating Eq.~\eqref{absorb} over the sample area. Then, the leading order becomes of the order of $O(R^2/\lambda^2)$. On the contrary, if we use conventional Gaussian beams, since they are approximately plane waves near the focus, the leading order is independent of the ratio $R/\lambda$. Hence, compared to Gaussian beams, the electric absorption of a focused CVB would be suppressed by the factor of $O(R^2/\lambda^2)$. 

To make the contrast with Gaussian beams better, we define an ``effective conductivity'' $\sigma'_{\rm eff}(\omega) = \frac{R^2}{\lambda^2}\sigma'(\omega)$. Then, the electric absorption of CVBs by a material with $\sigma'(\omega)$ is equivalent to the absorption of a simple Gaussian beam by the material with a reduced electric conductivity $\sigma'_{\rm eff}(\omega)$. This re-interpretation helps us compare CVBs with Gaussian beams. If the sample size is $1$ $\mu$m, and the wavelength is $300$ $\mu$m (1THz), the effective conductivity $\sigma'_{\rm eff}$ becomes $10^{-5}$ smaller than the original electric conductivity~\footnote{Since the wavelength of a light at 1 THz is much longer than the lattice constant, field absorption is locally the same for Gaussian beams and CVB. Their difference comes only from their field distributions.}.

Below we use the ratio of the two contributions in Eq.~\eqref{absorb}
\begin{align}\label{Pfactor}
P =  \frac{\omega \chi'' (\omega) B_{\rm CVB}^2}{ \sigma' (\omega) E_{\rm CVB}^2},
\end{align}
to discuss the relative strength of the magnetic field absorption. Here $B_{\rm CVB}$ and $E_{\rm CVB}$ are field amplitudes of the CVB. We ignore the $\rho$ component of the magnetic field and its absorption since it is geometrically suppressed just as $E^\phi$. By using the effective conductivity, we can rewrite Eq.~\eqref{Pfactor} as 
\begin{align}\label{Pfac_fin}
P =  \frac{\omega \chi'' (\omega) B_{\rm G}^2}{ \sigma'_{\rm eff} (\omega) E_{\rm G}^2} = \frac{\omega \chi'' (\omega)}{\sigma^{\rm eff}_0 c^2}
\end{align}
where $E_{\rm G}$ and $B_{\rm G}$ are field amplitudes of the corresponding Gaussian beams, and we used the relation $B_{\rm G} = E_{\rm G}/c$ to arrive at the final expression. For the order estimation, we also replace $\sigma'_{\rm eff}(\omega)$ by its DC value $\sigma^{\rm eff}_0$ since the electric conductivity in the THz region is almost the same as the DC value in most cases~\footnote{The Drude model of the electronic conduction predicts $\sigma'(\omega) = \sigma_0/(1 + \omega^2 \tau^2)$. The relaxation time of conduction electrons $\tau$ is typically order of femtosecond and thus the product $\omega \tau$ is, for $\omega$ in the THz region, quite small.}.

As an example, let us consider a two-dimensional (2D) electric conductor under a static in-plane magnetic field $H_0$ which induces a net  spin polarization of conduction elections. We apply a tightly focused azimuthal CVB from the out-of-plane direction and measure the ESR signals coming from the conduction electrons. To estimate the magnitude of the magnetic field absorption by conduction electrons, we use a phenomenological equation of motion of the magnetization dynamics, Bloch equation~\cite{PhysRev.104.563}. 

By solving the Bloch equation (see Appendix~\ref{app: bloch}), the imaginary part of the magnetic susceptibility is derived to be 
\begin{align} \label{chi}
 \chi'' \simeq \frac{\chi_0}{2} \gamma H_0 T_2 \frac{1}{1 + (\omega - \gamma H_0)^2 T_2^2}
\end{align}
where $T_2$ is the transverse relaxation time of the magnetization, and $\chi_0$ is the static susceptibility (real part of the susceptibility at $\omega = 0$). The resonance frequency and thus the energy of the Zeeman coupling with the external static magnetic field $H_0$ are assumed to be around THz. Namely, the external Zeeman field is assumed to be quite strong here, larger than ten Tesla. At the resonance $\omega = \gamma H_0 \equiv \omega_0$, we have $\chi '' =\chi_0(\omega_0 T_2)/2 $. If we take the transverse relaxation time to be about a nanosecond (typical timescale of the magnetization dynamics), the peak height will be $\chi'' \simeq 10^3\chi_0 /2$. Putting this expression in Eq.~\eqref{Pfac_fin}, in the SI unit, we obtain $P \sim 0.02 \frac{\chi_0}{\sigma^{\rm eff}_0} =  2000 \frac{\chi_0}{\sigma_0}.$ The dimensionless factor $P$ defines a criterion of whether the magnetic field absorption peaks are detectable or not. 

\begin{figure}[htbp]
\centering
\includegraphics[width = 80mm]{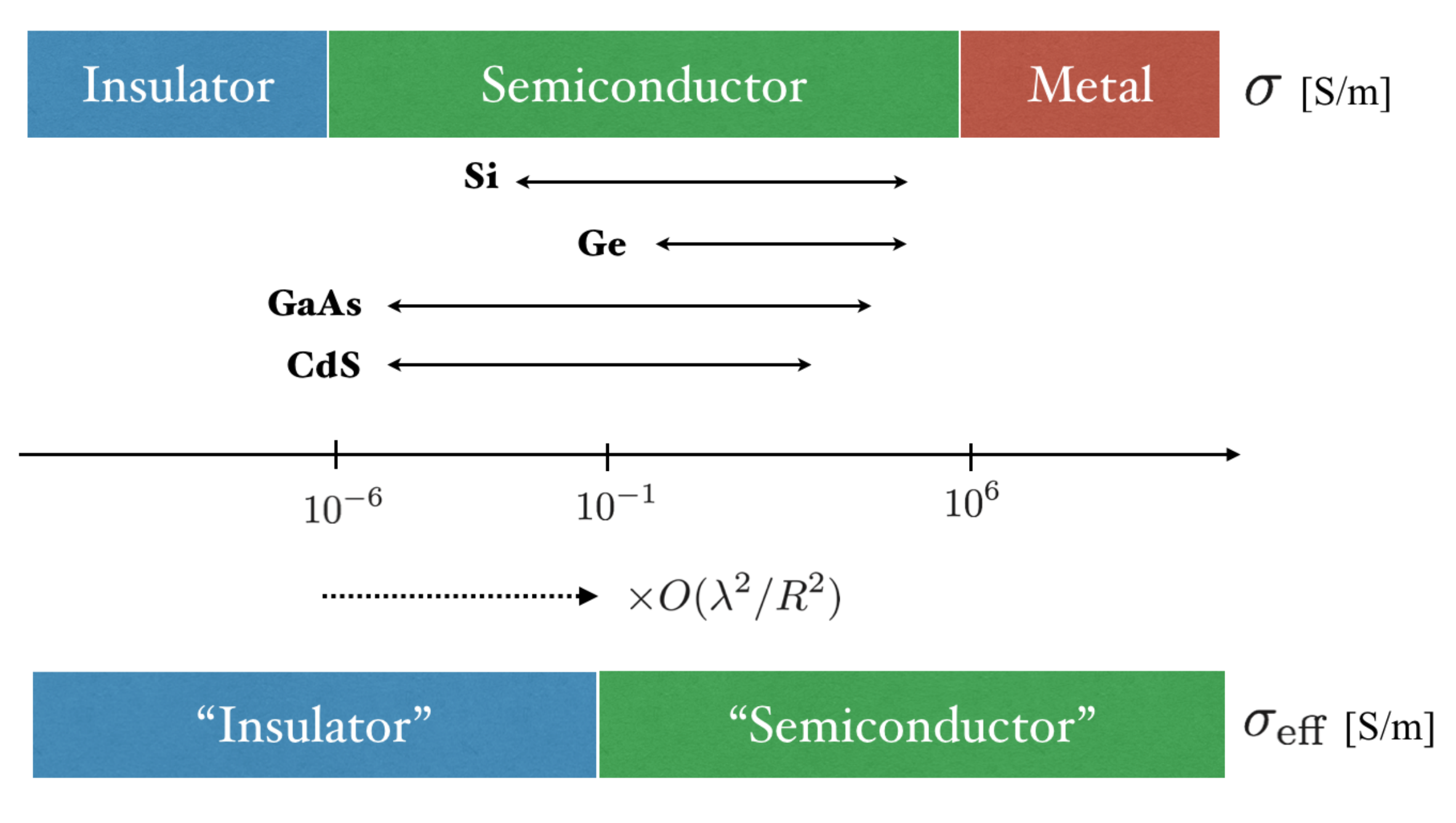}
\caption{Effective electric conductivity $\sigma_{\rm eff}$ for CVBs compared with the usual one $\sigma$ relevant for Gaussian beams. If the ratio of the wavelength and the system size satisfies $(R/\lambda)^2 = 10^{-5}$, various semiconductors including silicon, gallium arsenide, and cadmium sulfide become ``insulating'' in terms of the effective conductivity. We show the typical ranges of electric conductivity of several semiconductors taken from a literature~\cite{semicon_britannica}. }
       \label{cond}
  \end{figure}

In order for the position and line width of magnetic resonance peaks to be obtained, their height should be larger than noises or those from the conductivity microstructures of the electric field absorption. In the later section, we will give a quantitative argument specifically for EDSR absorption but here we give a rough estimate of the criterion, just assuming that the microstructure has the amplitude 1\% of $\sigma_0$ with no regard to its origin. Under this assumption, $P > 0.01$, namely $\sigma_0 < 2\times 10^5 \chi_0$ has to be satisfied. If $\chi_0 \sim 4\pi \times10^{-7}$[H$/$m], this gives the condition $\sigma_0 \le 0.3$ [S/m]. On the other hand, if we use the Gaussian beam in the first place, the same parameters lead to the criterion $\sigma_0 \le 0.3\times(R^2/\lambda^2) \sim 3\times10^{-6}$ [S/m]. The former can be satisfied in various semiconductors (see Fig.~\ref{cond}) but the latter is true only for insulators. By using CVBs, therefore, we could observe THz time-resolved ESR of conduction electrons in semiconductors. 

As is evident from the argument above, the criterion for the electric conductivity depends on the magnetic susceptibility and the magnitude of the conductivity microstructures. Since magnetic materials can have much larger magnetic susceptibility than that of a vacuum $\chi = 4\pi \times10^{-7}$[H$/$m], there would exist many magnetic semiconductors where CVBs become a powerful probe of their magnetic properties.

\subsection{Discussion}
The magnetic absorption does not necessarily come from conduction-electron spins. The magnetic degrees of freedom can be magnetic impurities, localized moments, and so on. Moreover, the origin of the magnetic excitation is not limited to the Zeeman splitting. In particular, it can be zero-field splitting due to magnetic anisotropy. As we will discuss later, in antiferromagnetic systems, the splitting from magnetic anisotropy is essential.

Although we have focused on electric absorption by conduction electrons in conductors,
the same argument applies to dielectric absorption by electric dipole moments in insulating systems like multiferroics or polar liquids. By using CVBs, we can suppress dielectric contribution to the electromagnetic field absorption. In the next subsection, we consider applications for both conducting and dielectric systems.

We again note that we have been considering pulse CVBs, having time-resolved measurements primarily in mind. If a powerful THz light source generating continuous waves is developed, we can use that to perform frequency-domain measurements. However, in this case CVBs may not be the best option. If continuous waves are available, we may use resonators to form standing waves and exploit their spatial property to suppress the electronic absorption. In the lower frequency region, this approach has been established~\cite{Barnes:1997aa,Reijerse2009}, and ESR measurements of heavy fermion metals~\cite{PhysRevLett.91.156401,PhysRevB.79.045104} are achieved at frequencies up to 360 GHz~\cite{Reijerse2009}.

\subsection{Applications}
As long as the sample size is small, the suppression of the electric field absorption for CVBs holds true for any systems. Below we provide several examples where focused pulse CVBs at (sub-) THz frequency would be useful.

\subsubsection{Magnetic resonance in conducting systems}
ESR of conduction electrons studied in Sec.~\ref{suppress} is a prototypical example where electric absorption is harmful. As we mentioned above, experimentally the magnetic absorption can be from either conduction electrons or localized magnetic moments. In any case, the success of the measurement relies on the relative magnitude of the magnetic absorption and the electric one corresponding to the conductivity microstructures . In this part, we give more detailed arguments on applications for magnetic resonance measurements. We firstly discuss the effect of SOC in more detail and then consider resonances in systems with large zero-field splittings.

EDSR is a primary source of the conductivity microstructures in electric conductors with SOC. Let us take a 2D electron system with a Rashba-type SOC~\cite{RevModPhys.87.1213} to make the situation more clear. The Hamiltonian of the SOC is given by
\begin{align}\label{SOC}
H_{\rm RSOC} \propto \sum_{\bm{p},\alpha,\beta} c_{\bm{p}, \alpha}^{\dagger} (\bm{d}_{\bm{p}}\cdot\bm{\sigma}_{\alpha, \beta}) c_{\bm{p},\beta},
\end{align}
where $\bm{\sigma}$ are Pauli matrices with matrix indices $\alpha$ and $\beta$ representing electron spins and $\bm{d}_{\bm{p}}$ is the SOC vector. The operator $c_{\bm{p}, \alpha}$ ($c^\dagger_{\bm{p}, \alpha}$) is the annihilation (creation) operator of electrons with a momentum $\bm{p}$ in the second quantization form~\cite{Ashcroft}. The SOC vector $\bm{d}_{\bm{p}}$ depends on the momentum of the electron and is taken to be, for example, $\bm{d}_{\bm{p}} \propto \hat{n}\times\bm{p}$ where $\hat{n}$ is the unit vector in the out-of-plane direction. As a result, the SOC term can be regarded as a Zeeman coupling with the momentum-dependent field and thus leads to the different spin-splitting for each (crystal) momentum. The momentum-induced splitting makes the system active to electric fields, resulting in the strong and broad EDSR absorption which often smears out the ESR contributions. 

The actual magnitude of the EDSR absorption strongly depends on the electron density of the system. To make the argument more quantitative, we consider the following one-dimensional tight-binding model
\begin{align}\label{Bolens}
    H = -t_0\sum_{i,\alpha} c^{\dagger}_{i+1, \alpha}c_{i, \alpha}+ h.c. - g\mu_B B\sum_{i,\alpha,\beta} c^{\dagger}_{i, \alpha}\sigma^{x}_{\alpha,\beta}c_{i, \beta} \nonumber \\ + i \lambda_R \sum_{i,\alpha,\beta} c^{\dagger}_{i+1, \alpha}\sigma^{y}_{\alpha,\beta}c_{i, \beta} + h.c.,
\end{align}
where the second term is the Zeeman coupling with the external field, and the third one is the Rashba SOC. Here $t_0$ is the electron hopping, $g$ is the g-factor, $\mu_B$ is the Bohr magneton, and $\lambda_R$ is the SOC constant. The operators $c_{i, \alpha}^{\dagger}$ and $c_{i, \alpha}$ create and annihilate electrons with spin $\alpha$ at the lattice site $i$. The EDSR absorption of the Hamiltonian~\eqref{Bolens} has been studied in Ref.~\onlinecite{PhysRevB.95.235115}.  
Although this is a one-dimensional system, we can consider a collection of such conducting chains and calculate the standard electric conductivity. 

We assume $\lambda_R/t_0 =  g\mu_B B/t_0 = 10^{-3}$ and take the intra- and inter- chain lattice constants to be $a_{\parallel} = a_{\perp} = 0.5$ nm.  As the electron hopping $t_0$ is typically order of electron volt, the magnitudes of $\lambda_R$ and $g\mu_B B$ are of THz. According to the calculation in Ref.~\onlinecite{PhysRevB.95.235115}, in this case, the EDSR contribution to electric conductivity for $\omega \sim g\mu_B B$ takes the range of $10^{-8}$ to $10^{-1}$ [S$/$m] depending on the electron density. Comparing this with Fig.~\ref{cond}, we notice that the EDSR contribution is in the ``insulating'' region of $\sigma_{\rm eff}$ relevant to CVBs. Therefore, by using CVBs, at least for a simple tight-binding model, absorption through the EDSR mechanism is well suppressed, making it possible for us to observe ESR signals even in the presence of EDSR. The estimated magnitude of the EDSR absorption is consistent with the value assumed in Sec.\ref{suppress}.

So far we have assumed simple magnetic resonance of paramagnetic or ferromagnetic systems whose resonance frequency is determined by the external static magnetic field. Hence, to lift the resonance frequency to be in the THz region, we have to apply a very large magnetic field over ten Tesla. On the other hand, in the case of conductors with localized moments, the resonance frequency can be very high even without such an external field. For example, in antiferromagnets, due to the mechanism so-called the exchange enhancement, the frequency of the antiferromagnetic resonance (AFMR) coming from a magnetic anisotropy is often lifted to the (sub-) THz region~\cite{Kampfrath:2011aa,PhysRevB.87.094422}.
Specifically, when the magnetic moments are coupled with each other through the antiferromagnetic exchange coupling $J_{\rm ex}$ and under the influence of the uniaxial anisotropy $A$, according to the Bloch equation for antiferromagnets, the resonance frequency is given by $\omega_{\rm res} = \sqrt{J_{\rm ex}A}$.

We note that in the case of paramagnetic or ferromagnetic conductors, the advantage of measurements using CVBs compared to the lower frequency ones for example with resonators~\cite{Reijerse2009} is apparent only in their temporal resolution. On the contrary, in the case of antiferromagnets, due to the large zero-field splitting, using high-frequency lights is inevitable to measure their magnetic resonances. Therefore, CVBs will have broad applications for studying AFMRs in the presence of electric absorption through, for instance, EDSR.

\subsubsection{Multiferroics}
As we noted, electric dipoles in insulators can be a source of electric absorption. In this and the next parts, we consider insulating systems with dielectric loss.

In a class of matter, multiple long-range orders appear at the same time, referred as multiferroics. Among multiferroics, matter with both ferromagnetic and ferroelectric orders is intensively studied. The magnetization and electric polarization in such materials mutually affect~\cite{0034-4885-77-7-076501,PhysRevLett.95.057205,PhysRevLett.104.177206,Pimenov:2006aa,PhysRevLett.98.027203,Takahashi:2012aa,PhysRevB.80.100402,PhysRevLett.105.257205} with each other, providing a way of controlling the ferroelectricity with a magnetic field and the magnetization with an electric field. As electric polarization in multiferroic materials depends on the configuration of localized magnetic moments, the elementary magnetic excitation of multiferroic materials, called the electromagnon, is active to both electric and magnetic components of the incident electromagnetic fields.

 If we use conventional Gaussian beams, both electric and magnetic fields are absorbed, and it is a nontrivial task to separate them out. If we know the magneto-electric coupling of the target beforehand, we could avoid the complication by combining measurements under different conditions. For example, we can use beams with different optical polarizations to identify the electric and magnetic contributions~\cite{Takahashi:2012aa}. However, this approach is not always works, and, in particular, is not applicable to newly synthesized materials with an unknown magneto-electric coupling.

As a proof of concept, let us take a simple toy model. We consider a pair of classical spins $\bm{S}_1$ and $\bm{S}_2$ under a linearly polarized beam $\bm{B}(t) = B \hat{y} \cos(\omega t)$, $\bm{E}(t) = E \hat{z} \sin(\omega t)$. We assume that the electric field couples with spins through the spin-current mechanism~\cite{PhysRevLett.95.057205}; $\bm{p} = \lambda_c \bm{e}_{1,2}\times(\bm{S}_1\times\bm{S}_2)$ where $\bm{e}_{1,2}=\hat{x}$ is the unit vector in the direction connecting these spins. 

We are going to study the laser-driven dynamics of the spin $\bm{S}_1(t)$. If there exists a strong static magnetic field in the $x$ direction, both spins stay almost fully-polarized in that direction. As long as we are interested in the small-amplitude dynamics of $\bm{S}_1$, therefore, we can approximately ignore that of $\bm{S}_2$ and take $\bm{S}_2 = \hat{x}$. Indeed, this corresponds to dropping out the higher-order terms of the dynamics of $\bm{S}_1$. In this case, the Hamiltonian for the spin $\bm{S}_1$ can be written as
\begin{align}
    H &= - \bm{p}\cdot\bm{E}(t) -g\mu_B \bm{S}_1\cdot\bm{B}(t)- g \mu_B H_0 S_1^x \nonumber \\
    &= -\lambda_c  S_1^z E\sin(\omega t)  - g \mu_B S_1^y B \cos(\omega t) - g\mu_B H_0 S_1^x,
\end{align}
Here, the static magnetic field $H_0$ incorporates the interaction between $\bm{S}_1$ and $\bm{S}_2$.

As shown in the appendix~\ref{eq10}, the time evolution of $S_1^y (t)$ is given as
\begin{align}\label{llg}
    S_1^y(t) = \frac{\gamma^2 H_0 B + \omega \lambda_c E}{\gamma^2 H_0^2 - \omega^2}\cos(\omega t)
\end{align}
where $\gamma = g\mu_B/\hbar$ is the gyromagnetic ratio. We see that the spin dynamics comes from both Zeeman and magneto-electric couplings with the beam. 

In the present case, as we know the functional form of the dynamics Eq.~\eqref{llg}, we can read out magnetic and electric contributions by measuring the time-evolution while changing parameters like $\omega$ and $H_0$. However, in general we do not even know the actual mechanism of the magneto-electric coupling so that examining parameter dependence is of little help for understanding the field-induced spin/polarization dynamics. With focused CVBs, we can independently apply the electric and magnetic field to the target. We can selectively activate the magnetization/polarization dynamics to study the effect of electric and magnetic fields without any prior knowledge of the spin-polarization coupling of the target. The use of CVBs for studying multiferroics will, therefore, largely streamline the characterization of such materials.

\subsubsection{Electron paramagnetic resonance in absorbing media}\label{sub:absorbint}

So far, we have been considering condensed-matter applications of CVBs. The potential use of CVBs is, however, not limited to solid-state physics. In this part, we give an example for chemistry and biology; a pulse electron paramagnetic resonance (EPR) of (bio) molecules in absorbing media.

EPR~\cite{Gunnar:2002aa,weil2007electron,halliwell2015free} is a primary tool of chemistry and biology where the absorption spectrum works as a fingerprint of unpaired electrons in the target. Specifically, in the spin label (or spin probe) method of EPR, we introduce a stable radical as a spin marker and measure its EPR absorption to study macromolecules like a protein.

In order to understand the dynamical properties of molecules, we have to resort to the time-dependent EPR using pulse waves. Since the temporal resolution of the measurement is determined by the pulse width, it is advantageous to use high-frequency fields for EPR. 

The problem is that when the target molecules are dissolved in absorbing media (e.g., liquid water), lightwave at (sub-) THz region is strongly absorbed by the environment, and thus the EPR of solute molecules becomes difficult~\cite{Nagy:2011aa}. In a biological or medical context, it is essential to study the nature of biological molecules in the living environment. Hence, the (sub-) THz absorption by an absorbing medium, especially liquid water, is a problem.

By using a pulse CVB, we can avoid exciting the oscillation modes of solvent polar molecules and perform the time-dependent EPR of solute molecules. Since the energy of spin triplet excitations of liquid water is high (of the order of eV or higher)~\cite{water}, the dominant magnetic absorption of the (sub-) THz CVBs would be, if exists, given by the solute molecules like spin markers of the spin label/probe methods. 

As a spin marker, commonly a nitroxide radical and molecules with that are used. The highly anisotropic nature of the $g$-factor and the hyperfine coupling constant of nitrogen allow high precision measurements and make nitroxides be the standard for EPR. However, if we are to use nitroxides as markers of the high-frequency pulse EPR, we have to apply a strong static magnetic field to lift the Zeeman splitting energy, making the measurement difficult. 

Another option for high-frequency pulse EPR is markers with large zero-field splitting like single-molecule magnets (SMM)~\cite{PhysRevB.56.8192,Edwards:2003aa,Yang:2005aa}. For example, if there are at least two magnetic ions in an SMM interacting with each other, the energy splitting among different spin multiplets naturally resides in the THz region. Even if an SMM contains only one magnetic ion, recently it is possible to design a molecular structure having a large magnetic anisotropy and thus a large zero-field splitting~\cite{doi:10.1021/ja308146e,C5SC02854J}. The use of CVBs, therefore, allows us to use a variety of SMMs as spin markers of high-frequency pulse EPR.

\section{Azimuthal electric field: imaging and controlling circulating currents}
So far, we have examined the use of the magnetic field component of the azimuthal CVBs. We discussed how their geometrical feature can be utilized to suppress the electric field absorption, and how to exploit that for measuring magnetic properties of matter. In this section, we focus on the {\it electric field component} of azimuthal CVBs which has been a nuisance in the previous section. As discussed in Sec.~\ref{sec:CVB}, the longitudinal magnetic field is prominent only if the focusing is tight so that for a weakly focused CVBs, electric absorption is dominant. In this section, we are only interested in the azimuthal nature of the electric field so that we assume a weakly focused CVB in the following argument and neglect the magnetic field component. 
The purely azimuthal electric field is written in the following form:
\begin{align}\label{aziE}
\bm{E}(\rho, \phi) = \frac{C}{\sqrt{w}} \left(\frac{\rho}{w}\right)\hat{e}_\phi e^{-i \omega t - \rho^2/w^2},
\end{align}
where $C$ is a constant determining the field amplitude and $\hat{e}_\phi$ is the unit vector in the azimuthal direction. This field indeed has the azimuthal polarization and is obtained by superimposing two optical vortices. The spatial profile is controlled by the beam width $w$, which should be larger than the wavelength because of the diffraction limit. We again note that unlike the previous sections, the arguments in this section are independent of the frequency/wavelength of the CVBs. In particular, the sample size has not to be small compared to the beam width in the following.

The azimuthal nature of the electric field of CVBs allows the field to couple with circulating currents efficiently. Here we consider its application to visualize and control the edge circulating current in topological insulators~\cite{RevModPhys.82.3045,RevModPhys.83.1057}. Topological insulators are characterized by the coexistence of the insulating bulk and metallic edges$/$surfaces. Notable examples of topological insulators in 2D are quantum Hall insulators and quantum spin Hall insulators. The former has metallic edges with chiral transport and realized for example in 2D electron systems under a strong out-of-plane magnetic field. The latter is schematically a combination of a pair of the former in a way restoring the time-reversal symmetry. The edge states in the quantum spin Hall insulators are thus helical consisting of two counter-propagating modes with the opposite spins. The metallic edges in those materials naturally host circulating electric current transport. Both the quantum Hall insulators and quantum spin Hall insulators are experimentally studied well. The existence of edge states has been verified by the transport measurements~\cite{Roth294} but its direct visualization is also important~\cite{Chen:2014aa,Susstrunk:2015aa,PhysRevLett.107.256803,PhysRevLett.107.176809,Nowack:2013aa}. Here we discuss how to use a CVB to visualize the edge states in topological insulators.

\subsection{Edge visualization}\label{visual}
Let us take a disk-shaped 2D topological insulator (quantum spin Hall insulator) with helical edge modes. We consider applying the azimuthal field Eq.~\eqref{aziE} and measure the electric absorption. The edge current can be modeled as a spatially inhomogeneous electric conductivity: $\sigma(\rho) \propto \Theta(R-\rho)e^{2\frac{\rho-R}{\xi}}$ where $R$ is the sample radius. The localization length $\xi$ is determined by the bulk property and typically of the order of nm. The function $\Theta(x)$ is the Heaviside Theta function. 

\begin{figure}[htbp]
\centering
\includegraphics[width = 80mm]{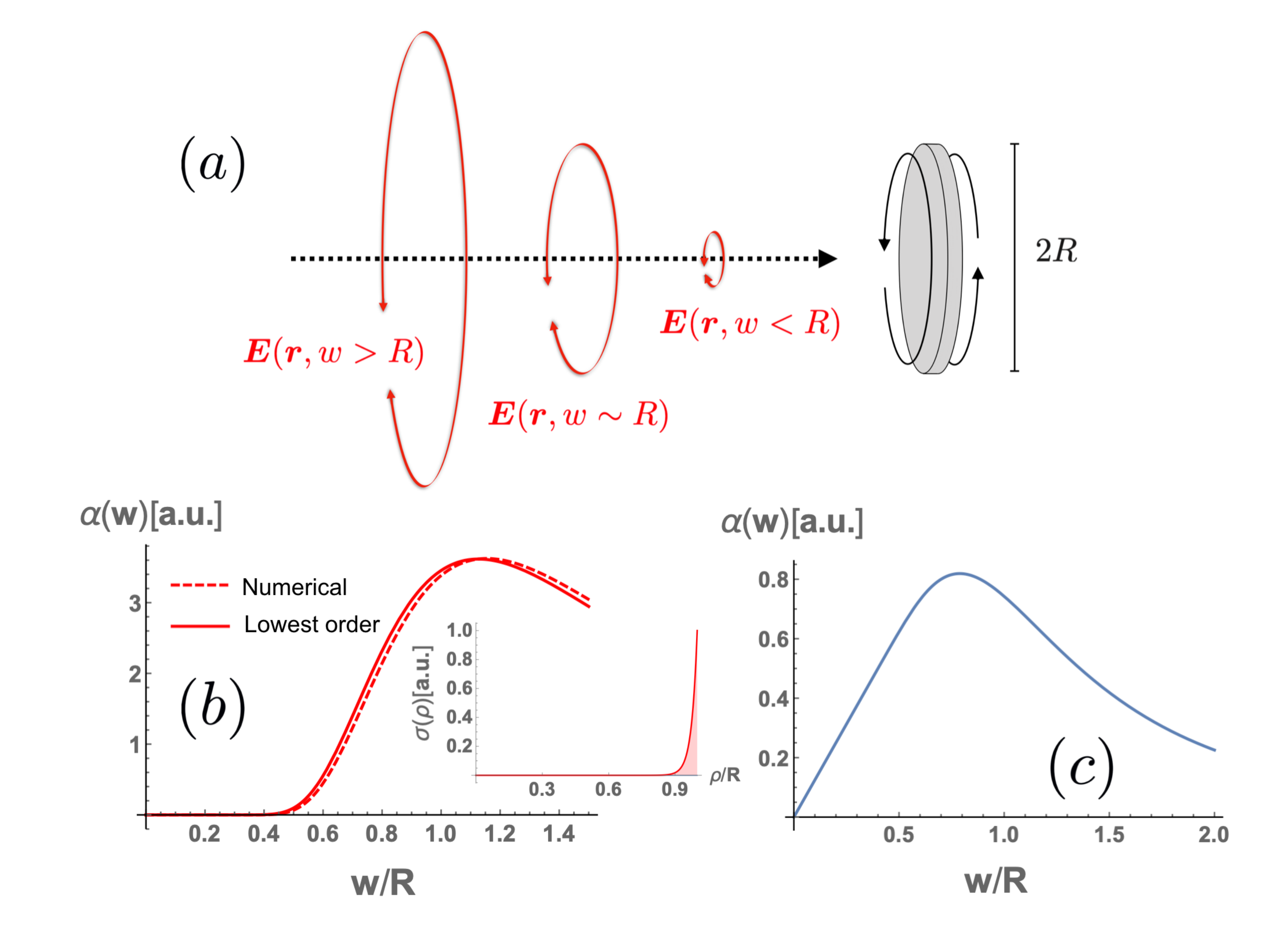}
\caption{Electric field absorption by edge currents. $(a)$ schematic illustration of the setup. Azimuthal electric fields with various beam widths are applied to a disk of a topological insulator with helical edge modes. $(b)$ Absorption as a function of the beam width $w$. We take the localization length $\xi = 0.05 R$. The inset shows the conductivity profile assumed in the calculation. $(c)$ Corresponding calculation for an ordinary metal with homogeneous conductivity distribution.}
       \label{integrated}
\end{figure}

The field amplitude of Eq.~\eqref{aziE} has its peak position determined by the beam width $w$. Then, we can expect that when the peak position matches with the position of the electric current, the electric absorption grows. That is, by measuring the electric absorption while changing the beam width, we can quantify where and how much the electric current is localized [see Fig.~\ref{integrated}$(a)$]. In the following, we examine this expectation for the edge localized currents. As a reference, we also consider the case with homogeneous conductivity which corresponds to ordinary metals.

The electric field absorption is, as we have seen in Eq.~\eqref{absorb}, given by
\begin{align}\label{abs_edge}
\alpha(w) &\propto \int_0^R \frac{1}{w} e^{2\frac{\rho-R}{\xi}}\left(\frac{\rho}{w}\right)^2e^{-2\rho^2/w^2}  \rho d\rho.
\end{align}
This integral can be analytically performed but the expression is lengthy and thus we omit that (see appendix B). In the lowest order of the localization length $\xi$, it simplifies to 
\begin{align}\label{simple}
   \alpha(w)\propto \xi  R^3 e^{-\frac{2 R^2}{w^2}}/(2 w^3).
\end{align}
The analytical expression will be useful as a fitting function in the experiment.

We show the $w$ dependence of the absorption in Fig.~\ref{integrated}$(b)$. The spatial profile of the conductivity for the localization length $\xi = 0.05 R$ is shown as the inset. We see that there exists a steep change at around $w = 0.5 R$, and the absorption takes its maximum when $w \sim R$ as we expected. Contrary to that, as shown in Fig.~\ref{integrated}$(c)$, if the target is an ordinary metal with spatially homogeneous conductivity $\sigma(\rho)  \propto  \Theta(R-\rho)$, there is no such steep change and the $w$ dependence is linear in the small $w$ region. Also, the absorption peaks for the width $w$ smaller than $R$. The linear dependence on $w$ for $w \ll R$ is easily derived from Eq.~\eqref{abs_edge} by taking $\xi \rightarrow \infty$ which corresponds to $\sigma(\rho) \propto  \Theta(R-\rho)$. By reconstructing the spatial profile of the electric conductivity, or that of the edge modes from the line shape of the $w$ dependence, we can visualize the edge transport in topological insulators.

 The visualization of edge states is interesting but is not conceptually new. In artificial systems like photonic or phononic crystals, the visualization of edge states has been well-established~\cite{Chen:2014aa,Susstrunk:2015aa}. Even in solids, there are several reports based on microscopes and SQUID techniques~\cite{PhysRevLett.107.256803,PhysRevLett.107.176809,Nowack:2013aa}. Those spatially resolved spectroscopies are indeed other options for visualizing the circulating currents.

\subsection{Orbital magnetization}  

 Here we shortly discuss that CVBs allow us to control dynamics of the edge states 
in addition to the visualization of them. 
Let us consider the case of generating a circulating current in a particular direction with a short-pulse azimuthal CVB.
In the setup, we can control/create an orbital magnetization arising from the current~\cite{0953-8984-22-12-123201, doi:10.1142/S0217979211058912}:
\begin{align}
M_{\rm orb}^{\rm edge} &\propto \int_0^R 2\pi \bm{r} \times \bm{j}^{\rm edge} \rho d\rho =  \int_0^R 2\pi \rho^2 \sigma(\rho) E^{\phi}(\rho) d\rho.
\end{align}
 That is, if the pulse width of the incident azimuthal beam is short enough (a half-cycle pulse is ideal), the electric field of that beam has the clear direction in which the circulating current is driven to induce net orbital magnetization.

The total magnetization measured experimentally is a sum of the orbital and spin magnetizations~\footnote{More precisely, the orbital magnetization is a sum of the surface and bulk contributions. The latter which we ignored comes from the microscopic circulating motions of electrons around each atom. Since these motions do not couple to the azimuthal electric field, the bulk contribution to the orbital magnetization would be negligible in the present case.}. Since we are considering a weakly focused CVBs, the longitudinal component of magnetic fields and the spin magnetization induced by that are negligible. Then, the orbital magnetization induced by the azimuthal electric field pulse is measurable just by looking at the out-of-plane component of the total magnetization for example with magneto-optical means. Hence, by using CVBs, we can control the magnetic property of the target material relying solely on the electric field.

We note that the laser-induced orbital magnetization does not require the system to be a topological insulator. In metals, the induced magnetization will be much stronger than that in topological insulators due to the large bulk electric conductivity. In particular, in spin-orbit-coupled systems, due to the coupling between the spin and orbital magnetization, it might be possible to control the spin magnetization through the laser-induced orbital magnetization. Further studies on their interplay is thus an interesting research direction.

\subsection{Discussion}
The candidate system for the experiment proposed above would be a thin film of 3D topological insulators such as Bi compounds and HgTe wells~\cite{RevModPhys.82.3045,RevModPhys.83.1057}. While the quantum Hall insulator is the simplest topological system, the chiral nature of the edge state would cause a complication. It would be indeed interesting to study the response of such chiral edge states for example in quantum Hall insulator or chiral topological superconductors to CVB pulses. The interplay of the chiral nature of the edge mode and the azimuthal electric field pulse would result in nonreciprocal responses which may be useful as a fingerprint of chiral edges. 

Although we did not discuss in detail, controlling circulating currents in metallic or superconducting rings would also be interesting. For example, by using the tightly focused CVB, we can insert magnetic fluxes in a metallic ring within the optical timescale and study its effect on electron transports such as Aharonov-Bohm effect. If we consider superconductors, with a CVB, we can control the electronic property of superconductors in the timescale much faster than the typical timescale of superconducting fluctuations. Such ultrafast dynamics of superconductors is a frontier of the field of superconductivity, and recently, interesting results like photo-induced superconductivity are reported~\cite{Fausti189,doi:10.1080/00107514.2017.1406623}. The use of CVBs allows us to control the magnetic flux and circulating current in an ultrafast way, offering new methods of controlling the nature of superconductivity.

\section{Application to Floquet engineering}
The highly controllable nature of CVBs offers an ideal playground for Floquet engineering~\cite{doi:10.1080/00018732.2015.1055918,RevModPhys.89.011004,kitamura-oka} where we drive a system periodically for example with lasers to realize desired nonequilibrium states. In this section, we take a toy model of quantum magnets to demonstrate the advantage of CVBs for Floquet engineering over conventional lasers. Since our argument relies on the Floquet theorem, we assume a continuous wave of CVB (or a pulse CVB with relatively long pulse duration) in this section.

 As we discussed in the previous sections, a tightly focused azimuthal (radial) CVB develops an almost purely longitudinal magnetic (electric) field component near the focus. Hence, superimposing radial and azimuthal CVBs offers a way of shining electric and magnetic field independently for systems smaller than the wavelength. There, we can freely change the relative angles, phases, and amplitudes of those fields. In the context of Floquet engineering, this property makes CVBs extremely powerful tools for designing nonequilibrium states of matter.

In Floquet engineering, we are interested in the nonequilibrium states of matter under a periodic drive. The time-evolution in a time-periodic system is described by a Hamiltonian $H(t) = \sum_m \hat{H}_m e^{i m t}$ where $t$ is time and $\hat{H}_m$ is the $m$-th Fourier component of $H(t)$. For the time-evolution operator $U(t_1, t_2) = \mathcal{T}_t \exp\left[i\int_{t_1}^{t_2} H(t) dt\right]$ with $\mathcal{T}_t$ being the time-ordering operator, we introduce a ``Floquet effective Hamiltonian'' $H_F$ as 
\begin{align}
    U(0,T) = e^{i H_F T},
\end{align}
where $T$ is the period of the external drive. That is, at each stroboscopic time $t = T, 2T, ...$ the system looks as if following the static Hamiltonian $H_F$.

 When the driving frequency is sufficiently high, so-called the Floquet-Magnus expansion gives a simple formula of the Floquet effective Hamiltonian:
\begin{align}\label{FM exp}
    H_F = \hat{H}_0 + \sum_{m>0} \frac{[\hat{H}_{-m}, \hat{H}_m]}{2m\omega}  + O\left(\frac{1}{\omega^2}\right).
\end{align}
This effective Hamiltonian enables us to predict the nonequilibrium dynamics under the drive in an intuitive way. For example, if we consider a Zeeman coupling between spins and a circularly polarized magnetic fields, Eq.~\eqref{FM exp} predicts an emergence of a large synthetic magnetic field perpendicular to the propagating axis of the beam~\cite{slichter1996principles,PhysRevB.90.085150,PhysRevB.90.214413}, and this is consistent with the known inverse-Faraday effect. 

 Applications of Floquet engineering are widespread. For instance, it plays a central role in modern cold-atom physics as it allows us to synthesize effective electromagnetic fields and spin-orbit coupling for charge-neutral atoms~\cite{PhysRevX.4.031027}. In the context of condensed-matter physics, lasers are the most common source of the required periodic perturbation, and laser-induced topological phase transition in graphene was predicted~\cite{PhysRevB.79.081406} and experimentally verified~\cite{Wang453,arXiv:1811.03522}.

Let us take a simple example, laser-induced multiferroicity~\cite{arXiv:1602.03702} to show why Floquet engineering with CVBs is advantageous over that with other means like circularly polarized beams. We consider a pair of quantum spins $\bm{S}_1$ and $\bm{S}_2$ placed along the $x$ axis and apply CVBs to them. The spins couple to magnetic fields through the Zeeman coupling $H_B = -g \mu_B (\bm{S}_1 + \bm{S}_2)\cdot\bm{B}(t)  = g\mu_B B\cos(\omega t + \delta) (S^z_1 + S_2^z)$ where we consider a focused azimuthal CVB applied along the $z$ axis; $\bm{B}(t) = B\cos(\omega t + \delta) \hat{z}$. We assume that spins couple to electric fields through the magneto-electric coupling of the spin-current mechanism $H_p = -g_{me}\bm{p}\cdot\bm{E}$ with $\bm{p} = \hat{x}\times (\bm{S}_1\times\bm{S}_2)$, and take the electric field as $\bm{E} = E(\cos{\theta}\hat{z} + \sin{\theta}\hat{y})\cos(\omega t)$. That is, we apply a focused radial CVB with frequency $\omega$ along the direction $(0, \sin\theta,\cos\theta)$.  If we take $\delta = \pi/2$ and $\theta = \pi/2$, the field configuration is equivalent to a linearly polarized laser propagating in the $x$ direction. In our setup,  we can freely control $\delta$ and $\theta$ by changing the relative phase and the relative angle between the optical axes of the radial and azimuthal beams.

The Floquet effective Hamiltonian is obtained from Eq.~\eqref{FM exp} as 
\begin{align}\label{synth}
    H_F = \hat{H}_0 + \frac{\sin\delta}{4 \omega}gg_{me}\mu_B E B \cos{\theta} (\bm{S}_1\times\bm{S}_2)^x + O\left(\frac{1}{\omega^2}\right).
\end{align}
That is, the periodic drive by the focused CVBs results in the synthetic Dzyaloshinskii-Moriya (DM) type interaction~\cite{DZYALOSHINSKY1958241,PhysRev.120.91} between the spins. 
In the appendix D, we show that this Floquet effective Hamiltonian indeed gives an approximate description of the spin dynamics.

The leading order term of the Floquet Hamiltonian~\eqref{synth} contains only the synthetic DM interaction. This is in stark contrast to the result of the previous work~\cite{arXiv:1602.03702} dealing with the same model but using circularly polarized lasers applied along the $z$ axis instead. There, in addition to the synthetic DM interaction, we inevitably have a synthetic Zeeman field which is undesirable for exploiting the synthetic DM interaction for designing nonequilibrium states. 

We see that in Eq.~\eqref{synth}, there are a number of controllable parameters for CVBs; relative phase $\delta$, relative angle $\theta$, field amplitudes $E$ and $B$, frequency $\omega$. Such high controllability is achieved by the unique geometrical feature of focused CVBs. Since the typical timescale of a spin system is slower than THz, the high-frequency expansion is supposed to be a good approximation for the laser-driven dynamics. Therefore, using focused CVBs for Floquet engineering is a promising application.

\section{Conclusion}
In this paper, we explored the use of the (azimuthal) cylindrical vector beam for studying and controlling electromagnetic properties of matter. Compared to the conventional Gaussian beams, the electric field contribution to optical absorption is strongly suppressed for small samples. 

In the third section, we enumerated some applications; studies of magnetic resonances in conductors, simplified characterization of multiferroics, and high-frequency electron spin resonance of water solutions. Other examples may include the study of topological insulators, magnetic superconductors, and quantum dots/wells. Like the electron paramagnetic resonance in polar medium for biological purposes, the proposed spectroscopy using cylindrical vector beams has various applications even outside condensed matter physics. The high temporal resolution of THz time-domain spectroscopy with cylindrical vector beams would be a window to nonequilibrium magnetic properties of matter.

In the fourth section, we discussed the applications of the azimuthal electric field as well. It can couple to circulating currents and dynamically generate orbital magnetization. We could use those properties to study circulating edge modes in (topological) materials. It is an open question whether the photo-induced orbital magnetization has any applications in the field of spintronics. Also, it would be interesting to use cylindrical vector beams for controlling other types of circulating currents like those in superconducting rings or mesoscopic systems.

In the fifth section, we discussed Floquet engineering. Taking a simple model of a driven spin, we show that using focused cylindrical vector beams has clear advantages over known schemes. The extremely high controllability of focused cylindrical vector beams would offer building blocks of artificially designed nonequilibrium states of matter.

The spatial properties of cylindrical vector beams have various applications. We can use them for suppressing electric field absorption, controlling circulating currents in matter, and designing electromagnetic fields. The requirement on the sample size can be easily satisfied for various condensed matter and chemistry uses. Although THz optics is still a developing field, the proposed applications are feasible with current THz/far-infrared technologies. We expect that cylindrical vector beams would be powerful means for characterizing and controlling electromagnetic properties of matter at THz frequency.

\section{Acknowledgment}
We thank Shunsuke Furuya for useful comments. H. F. is supported by Advanced Leading Graduate Course for Photon Science (ALPS) of Japan Society for the Promotion of Science (JSPS) and JSPS KAKENHI Grant-in-Aid for JSPS Fellows Grant No.~JP16J04752. Y. T. is supported by Grants-in-Aid for Scientific Research No.~JP17K14333 and KAKENHI on Innovative Areas ``J-Physics'' [No. JP18H04318].  M. S. was supported by Grant-in-Aid for Scientific Research on Innovative Area, ``Nano Spin Conversion Science'' (Grant No.17H05174), and JSPS KAKENHI (Grant No. JP17K05513 and No. JP15H02117).

\clearpage

\onecolumngrid

\bibliography{CVspectroscopy}

\begin{thebibliography}{104}%
\makeatletter
\providecommand \@ifxundefined [1]{%
 \@ifx{#1\undefined}
}%
\providecommand \@ifnum [1]{%
 \ifnum #1\expandafter \@firstoftwo
 \else \expandafter \@secondoftwo
 \fi
}%
\providecommand \@ifx [1]{%
 \ifx #1\expandafter \@firstoftwo
 \else \expandafter \@secondoftwo
 \fi
}%
\providecommand \natexlab [1]{#1}%
\providecommand \enquote  [1]{``#1''}%
\providecommand \bibnamefont  [1]{#1}%
\providecommand \bibfnamefont [1]{#1}%
\providecommand \citenamefont [1]{#1}%
\providecommand \href@noop [0]{\@secondoftwo}%
\providecommand \href [0]{\begingroup \@sanitize@url \@href}%
\providecommand \@href[1]{\@@startlink{#1}\@@href}%
\providecommand \@@href[1]{\endgroup#1\@@endlink}%
\providecommand \@sanitize@url [0]{\catcode `\\12\catcode `\$12\catcode
  `\&12\catcode `\#12\catcode `\^12\catcode `\_12\catcode `\%12\relax}%
\providecommand \@@startlink[1]{}%
\providecommand \@@endlink[0]{}%
\providecommand \url  [0]{\begingroup\@sanitize@url \@url }%
\providecommand \@url [1]{\endgroup\@href {#1}{\urlprefix }}%
\providecommand \urlprefix  [0]{URL }%
\providecommand \Eprint [0]{\href }%
\providecommand \doibase [0]{http://dx.doi.org/}%
\providecommand \selectlanguage [0]{\@gobble}%
\providecommand \bibinfo  [0]{\@secondoftwo}%
\providecommand \bibfield  [0]{\@secondoftwo}%
\providecommand \translation [1]{[#1]}%
\providecommand \BibitemOpen [0]{}%
\providecommand \bibitemStop [0]{}%
\providecommand \bibitemNoStop [0]{.\EOS\space}%
\providecommand \EOS [0]{\spacefactor3000\relax}%
\providecommand \BibitemShut  [1]{\csname bibitem#1\endcsname}%
\let\auto@bib@innerbib\@empty
\bibitem [{\citenamefont {Strickland}\ and\ \citenamefont
  {Mourou}(1985)}]{STRICKLAND1985219}%
  \BibitemOpen
  \bibfield  {author} {\bibinfo {author} {\bibfnamefont {D.}~\bibnamefont
  {Strickland}}\ and\ \bibinfo {author} {\bibfnamefont {G.}~\bibnamefont
  {Mourou}},\ }\href {\doibase https://doi.org/10.1016/0030-4018(85)90120-8}
  {\bibfield  {journal} {\bibinfo  {journal} {Optics Communications}\ }\textbf
  {\bibinfo {volume} {56}},\ \bibinfo {pages} {219 } (\bibinfo {year}
  {1985})}\BibitemShut {NoStop}%
\bibitem [{\citenamefont {Backus}\ \emph {et~al.}(1998)\citenamefont {Backus},
  \citenamefont {Durfee}, \citenamefont {Murnane},\ and\ \citenamefont
  {Kapteyn}}]{doi:10.1063/1.1148795}%
  \BibitemOpen
  \bibfield  {author} {\bibinfo {author} {\bibfnamefont {S.}~\bibnamefont
  {Backus}}, \bibinfo {author} {\bibfnamefont {C.~G.}\ \bibnamefont {Durfee}},
  \bibinfo {author} {\bibfnamefont {M.~M.}\ \bibnamefont {Murnane}}, \ and\
  \bibinfo {author} {\bibfnamefont {H.~C.}\ \bibnamefont {Kapteyn}},\ }\href
  {\doibase 10.1063/1.1148795} {\bibfield  {journal} {\bibinfo  {journal}
  {Review of Scientific Instruments}\ }\textbf {\bibinfo {volume} {69}},\
  \bibinfo {pages} {1207} (\bibinfo {year} {1998})}\BibitemShut {NoStop}%
\bibitem [{\citenamefont {Nasu}(2004)}]{doi:10.1142/5476}%
  \BibitemOpen
  \bibfield  {author} {\bibinfo {author} {\bibfnamefont {K.}~\bibnamefont
  {Nasu}},\ }\href {\doibase 10.1142/5476} {\emph {\bibinfo {title}
  {Photoinduced Phase Transitions}}}\ (\bibinfo  {publisher} {WORLD
  SCIENTIFIC},\ \bibinfo {year} {2004})\BibitemShut {NoStop}%
\bibitem [{\citenamefont {Kirilyuk}\ \emph {et~al.}(2010)\citenamefont
  {Kirilyuk}, \citenamefont {Kimel},\ and\ \citenamefont
  {Rasing}}]{RevModPhys.82.2731}%
  \BibitemOpen
  \bibfield  {author} {\bibinfo {author} {\bibfnamefont {A.}~\bibnamefont
  {Kirilyuk}}, \bibinfo {author} {\bibfnamefont {A.~V.}\ \bibnamefont {Kimel}},
  \ and\ \bibinfo {author} {\bibfnamefont {T.}~\bibnamefont {Rasing}},\ }\href
  {\doibase 10.1103/RevModPhys.82.2731} {\bibfield  {journal} {\bibinfo
  {journal} {Rev. Mod. Phys.}\ }\textbf {\bibinfo {volume} {82}},\ \bibinfo
  {pages} {2731} (\bibinfo {year} {2010})}\BibitemShut {NoStop}%
\bibitem [{\citenamefont {Zewail}(1988)}]{Zewail1645}%
  \BibitemOpen
  \bibfield  {author} {\bibinfo {author} {\bibfnamefont {A.~H.}\ \bibnamefont
  {Zewail}},\ }\href {\doibase 10.1126/science.242.4886.1645} {\bibfield
  {journal} {\bibinfo  {journal} {Science}\ }\textbf {\bibinfo {volume}
  {242}},\ \bibinfo {pages} {1645} (\bibinfo {year} {1988})}\BibitemShut
  {NoStop}%
\bibitem [{\citenamefont {Tonouchi}(2007)}]{Tonouchi:2007aa}%
  \BibitemOpen
  \bibfield  {author} {\bibinfo {author} {\bibfnamefont {M.}~\bibnamefont
  {Tonouchi}},\ }\href {http://dx.doi.org/10.1038/nphoton.2007.3} {\bibfield
  {journal} {\bibinfo  {journal} {Nature Photonics}\ }\textbf {\bibinfo
  {volume} {1}},\ \bibinfo {pages} {97 EP } (\bibinfo {year}
  {2007})}\BibitemShut {NoStop}%
\bibitem [{\citenamefont {George}\ \emph {et~al.}(2008)\citenamefont {George},
  \citenamefont {Strait}, \citenamefont {Dawlaty}, \citenamefont {Shivaraman},
  \citenamefont {Chandrashekhar}, \citenamefont {Rana},\ and\ \citenamefont
  {Spencer}}]{George:2008aa}%
  \BibitemOpen
  \bibfield  {author} {\bibinfo {author} {\bibfnamefont {P.~A.}\ \bibnamefont
  {George}}, \bibinfo {author} {\bibfnamefont {J.}~\bibnamefont {Strait}},
  \bibinfo {author} {\bibfnamefont {J.}~\bibnamefont {Dawlaty}}, \bibinfo
  {author} {\bibfnamefont {S.}~\bibnamefont {Shivaraman}}, \bibinfo {author}
  {\bibfnamefont {M.}~\bibnamefont {Chandrashekhar}}, \bibinfo {author}
  {\bibfnamefont {F.}~\bibnamefont {Rana}}, \ and\ \bibinfo {author}
  {\bibfnamefont {M.~G.}\ \bibnamefont {Spencer}},\ }\bibfield  {booktitle}
  {\emph {\bibinfo {booktitle} {Nano Letters}},\ }\href {\doibase
  10.1021/nl8019399} {\bibfield  {journal} {\bibinfo  {journal} {Nano Letters}\
  }\textbf {\bibinfo {volume} {8}},\ \bibinfo {pages} {4248} (\bibinfo {year}
  {2008})}\BibitemShut {NoStop}%
\bibitem [{\citenamefont {Liu}\ \emph {et~al.}(2012)\citenamefont {Liu},
  \citenamefont {Hwang}, \citenamefont {Tao}, \citenamefont {Strikwerda},
  \citenamefont {Fan}, \citenamefont {Keiser}, \citenamefont {Sternbach},
  \citenamefont {West}, \citenamefont {Kittiwatanakul}, \citenamefont {Lu},
  \citenamefont {Wolf}, \citenamefont {Omenetto}, \citenamefont {Zhang},
  \citenamefont {Nelson},\ and\ \citenamefont {Averitt}}]{Liu:2012aa}%
  \BibitemOpen
  \bibfield  {author} {\bibinfo {author} {\bibfnamefont {M.}~\bibnamefont
  {Liu}}, \bibinfo {author} {\bibfnamefont {H.~Y.}\ \bibnamefont {Hwang}},
  \bibinfo {author} {\bibfnamefont {H.}~\bibnamefont {Tao}}, \bibinfo {author}
  {\bibfnamefont {A.~C.}\ \bibnamefont {Strikwerda}}, \bibinfo {author}
  {\bibfnamefont {K.}~\bibnamefont {Fan}}, \bibinfo {author} {\bibfnamefont
  {G.~R.}\ \bibnamefont {Keiser}}, \bibinfo {author} {\bibfnamefont {A.~J.}\
  \bibnamefont {Sternbach}}, \bibinfo {author} {\bibfnamefont {K.~G.}\
  \bibnamefont {West}}, \bibinfo {author} {\bibfnamefont {S.}~\bibnamefont
  {Kittiwatanakul}}, \bibinfo {author} {\bibfnamefont {J.}~\bibnamefont {Lu}},
  \bibinfo {author} {\bibfnamefont {S.~A.}\ \bibnamefont {Wolf}}, \bibinfo
  {author} {\bibfnamefont {F.~G.}\ \bibnamefont {Omenetto}}, \bibinfo {author}
  {\bibfnamefont {X.}~\bibnamefont {Zhang}}, \bibinfo {author} {\bibfnamefont
  {K.~A.}\ \bibnamefont {Nelson}}, \ and\ \bibinfo {author} {\bibfnamefont
  {R.~D.}\ \bibnamefont {Averitt}},\ }\href
  {http://dx.doi.org/10.1038/nature11231} {\bibfield  {journal} {\bibinfo
  {journal} {Nature}\ }\textbf {\bibinfo {volume} {487}},\ \bibinfo {pages}
  {345 EP } (\bibinfo {year} {2012})}\BibitemShut {NoStop}%
\bibitem [{\citenamefont {Matsunaga}\ \emph {et~al.}(2013)\citenamefont
  {Matsunaga}, \citenamefont {Hamada}, \citenamefont {Makise}, \citenamefont
  {Uzawa}, \citenamefont {Terai}, \citenamefont {Wang},\ and\ \citenamefont
  {Shimano}}]{PhysRevLett.111.057002}%
  \BibitemOpen
  \bibfield  {author} {\bibinfo {author} {\bibfnamefont {R.}~\bibnamefont
  {Matsunaga}}, \bibinfo {author} {\bibfnamefont {Y.~I.}\ \bibnamefont
  {Hamada}}, \bibinfo {author} {\bibfnamefont {K.}~\bibnamefont {Makise}},
  \bibinfo {author} {\bibfnamefont {Y.}~\bibnamefont {Uzawa}}, \bibinfo
  {author} {\bibfnamefont {H.}~\bibnamefont {Terai}}, \bibinfo {author}
  {\bibfnamefont {Z.}~\bibnamefont {Wang}}, \ and\ \bibinfo {author}
  {\bibfnamefont {R.}~\bibnamefont {Shimano}},\ }\href {\doibase
  10.1103/PhysRevLett.111.057002} {\bibfield  {journal} {\bibinfo  {journal}
  {Phys. Rev. Lett.}\ }\textbf {\bibinfo {volume} {111}},\ \bibinfo {pages}
  {057002} (\bibinfo {year} {2013})}\BibitemShut {NoStop}%
\bibitem [{\citenamefont {Matsunaga}\ \emph {et~al.}(2014)\citenamefont
  {Matsunaga}, \citenamefont {Tsuji}, \citenamefont {Fujita}, \citenamefont
  {Sugioka}, \citenamefont {Makise}, \citenamefont {Uzawa}, \citenamefont
  {Terai}, \citenamefont {Wang}, \citenamefont {Aoki},\ and\ \citenamefont
  {Shimano}}]{Matsunaga1145}%
  \BibitemOpen
  \bibfield  {author} {\bibinfo {author} {\bibfnamefont {R.}~\bibnamefont
  {Matsunaga}}, \bibinfo {author} {\bibfnamefont {N.}~\bibnamefont {Tsuji}},
  \bibinfo {author} {\bibfnamefont {H.}~\bibnamefont {Fujita}}, \bibinfo
  {author} {\bibfnamefont {A.}~\bibnamefont {Sugioka}}, \bibinfo {author}
  {\bibfnamefont {K.}~\bibnamefont {Makise}}, \bibinfo {author} {\bibfnamefont
  {Y.}~\bibnamefont {Uzawa}}, \bibinfo {author} {\bibfnamefont
  {H.}~\bibnamefont {Terai}}, \bibinfo {author} {\bibfnamefont
  {Z.}~\bibnamefont {Wang}}, \bibinfo {author} {\bibfnamefont {H.}~\bibnamefont
  {Aoki}}, \ and\ \bibinfo {author} {\bibfnamefont {R.}~\bibnamefont
  {Shimano}},\ }\href {\doibase 10.1126/science.1254697} {\bibfield  {journal}
  {\bibinfo  {journal} {Science}\ }\textbf {\bibinfo {volume} {345}},\ \bibinfo
  {pages} {1145} (\bibinfo {year} {2014})}\BibitemShut {NoStop}%
\bibitem [{\citenamefont {Markelz}\ \emph {et~al.}(2000)\citenamefont
  {Markelz}, \citenamefont {Roitberg},\ and\ \citenamefont
  {Heilweil}}]{Markelz:2000aa}%
  \BibitemOpen
  \bibfield  {author} {\bibinfo {author} {\bibfnamefont {A.~G.}\ \bibnamefont
  {Markelz}}, \bibinfo {author} {\bibfnamefont {A.}~\bibnamefont {Roitberg}}, \
  and\ \bibinfo {author} {\bibfnamefont {E.~J.}\ \bibnamefont {Heilweil}},\
  }\href {\doibase https://doi.org/10.1016/S0009-2614(00)00227-X} {\bibfield
  {journal} {\bibinfo  {journal} {Chemical Physics Letters}\ }\textbf {\bibinfo
  {volume} {320}},\ \bibinfo {pages} {42} (\bibinfo {year} {2000})}\BibitemShut
  {NoStop}%
\bibitem [{\citenamefont {Fischer}\ \emph {et~al.}(2002)\citenamefont
  {Fischer}, \citenamefont {Walther},\ and\ \citenamefont
  {Jepsen}}]{0031-9155-47-21-319}%
  \BibitemOpen
  \bibfield  {author} {\bibinfo {author} {\bibfnamefont {B.~M.}\ \bibnamefont
  {Fischer}}, \bibinfo {author} {\bibfnamefont {M.}~\bibnamefont {Walther}}, \
  and\ \bibinfo {author} {\bibfnamefont {P.~U.}\ \bibnamefont {Jepsen}},\
  }\href {http://stacks.iop.org/0031-9155/47/i=21/a=319} {\bibfield  {journal}
  {\bibinfo  {journal} {Physics in Medicine and Biology}\ }\textbf {\bibinfo
  {volume} {47}},\ \bibinfo {pages} {3807} (\bibinfo {year}
  {2002})}\BibitemShut {NoStop}%
\bibitem [{\citenamefont {Markelz}\ \emph {et~al.}(2002)\citenamefont
  {Markelz}, \citenamefont {Whitmire}, \citenamefont {Hillebrecht},\ and\
  \citenamefont {Birge}}]{0031-9155-47-21-318}%
  \BibitemOpen
  \bibfield  {author} {\bibinfo {author} {\bibfnamefont {A.}~\bibnamefont
  {Markelz}}, \bibinfo {author} {\bibfnamefont {S.}~\bibnamefont {Whitmire}},
  \bibinfo {author} {\bibfnamefont {J.}~\bibnamefont {Hillebrecht}}, \ and\
  \bibinfo {author} {\bibfnamefont {R.}~\bibnamefont {Birge}},\ }\href
  {http://stacks.iop.org/0031-9155/47/i=21/a=318} {\bibfield  {journal}
  {\bibinfo  {journal} {Physics in Medicine and Biology}\ }\textbf {\bibinfo
  {volume} {47}},\ \bibinfo {pages} {3797} (\bibinfo {year}
  {2002})}\BibitemShut {NoStop}%
\bibitem [{\citenamefont {Plusquellic}\ \emph {et~al.}(2007)\citenamefont
  {Plusquellic}, \citenamefont {Siegrist}, \citenamefont {Heilweil},\ and\
  \citenamefont {Esenturk}}]{F.:2007aa}%
  \BibitemOpen
  \bibfield  {author} {\bibinfo {author} {\bibfnamefont {D.~F.}\ \bibnamefont
  {Plusquellic}}, \bibinfo {author} {\bibfnamefont {K.}~\bibnamefont
  {Siegrist}}, \bibinfo {author} {\bibfnamefont {E.~J.}\ \bibnamefont
  {Heilweil}}, \ and\ \bibinfo {author} {\bibfnamefont {O.}~\bibnamefont
  {Esenturk}},\ }\href {\doibase doi:10.1002/cphc.200700332} {\bibfield
  {journal} {\bibinfo  {journal} {ChemPhysChem}\ }\textbf {\bibinfo {volume}
  {8}},\ \bibinfo {pages} {2412} (\bibinfo {year} {2007})}\BibitemShut
  {NoStop}%
\bibitem [{\citenamefont {Youngworth}\ and\ \citenamefont
  {Brown}(2000)}]{Youngworth:00}%
  \BibitemOpen
  \bibfield  {author} {\bibinfo {author} {\bibfnamefont {K.~S.}\ \bibnamefont
  {Youngworth}}\ and\ \bibinfo {author} {\bibfnamefont {T.~G.}\ \bibnamefont
  {Brown}},\ }\href {\doibase 10.1364/OE.7.000077} {\bibfield  {journal}
  {\bibinfo  {journal} {Opt. Express}\ }\textbf {\bibinfo {volume} {7}},\
  \bibinfo {pages} {77} (\bibinfo {year} {2000})}\BibitemShut {NoStop}%
\bibitem [{\citenamefont {Zhan}\ and\ \citenamefont {Leger}(2002)}]{Zhan:02}%
  \BibitemOpen
  \bibfield  {author} {\bibinfo {author} {\bibfnamefont {Q.}~\bibnamefont
  {Zhan}}\ and\ \bibinfo {author} {\bibfnamefont {J.~R.}\ \bibnamefont
  {Leger}},\ }\href {\doibase 10.1364/OE.10.000324} {\bibfield  {journal}
  {\bibinfo  {journal} {Opt. Express}\ }\textbf {\bibinfo {volume} {10}},\
  \bibinfo {pages} {324} (\bibinfo {year} {2002})}\BibitemShut {NoStop}%
\bibitem [{\citenamefont {Zhan}(2009)}]{Zhan:09}%
  \BibitemOpen
  \bibfield  {author} {\bibinfo {author} {\bibfnamefont {Q.}~\bibnamefont
  {Zhan}},\ }\href {\doibase 10.1364/AOP.1.000001} {\bibfield  {journal}
  {\bibinfo  {journal} {Adv. Opt. Photon.}\ }\textbf {\bibinfo {volume} {1}},\
  \bibinfo {pages} {1} (\bibinfo {year} {2009})}\BibitemShut {NoStop}%
\bibitem [{\citenamefont {Allen}\ \emph {et~al.}(1992)\citenamefont {Allen},
  \citenamefont {Beijersbergen}, \citenamefont {Spreeuw},\ and\ \citenamefont
  {Woerdman}}]{PhysRevA.45.8185}%
  \BibitemOpen
  \bibfield  {author} {\bibinfo {author} {\bibfnamefont {L.}~\bibnamefont
  {Allen}}, \bibinfo {author} {\bibfnamefont {M.~W.}\ \bibnamefont
  {Beijersbergen}}, \bibinfo {author} {\bibfnamefont {R.~J.~C.}\ \bibnamefont
  {Spreeuw}}, \ and\ \bibinfo {author} {\bibfnamefont {J.~P.}\ \bibnamefont
  {Woerdman}},\ }\href {\doibase 10.1103/PhysRevA.45.8185} {\bibfield
  {journal} {\bibinfo  {journal} {Phys. Rev. A}\ }\textbf {\bibinfo {volume}
  {45}},\ \bibinfo {pages} {8185} (\bibinfo {year} {1992})}\BibitemShut
  {NoStop}%
\bibitem [{\citenamefont {Andrews}\ and\ \citenamefont
  {Babiker}(2012)}]{9780511795213}%
  \BibitemOpen
  \bibinfo {editor} {\bibfnamefont {D.~L.}\ \bibnamefont {Andrews}}\ and\
  \bibinfo {editor} {\bibfnamefont {M.}~\bibnamefont {Babiker}},\ eds.,\ \href
  {http://dx.doi.org/10.1017/CBO9780511795213} {\emph {\bibinfo {title} {{The
  Angular Momentum of Light}}}}\ (\bibinfo  {publisher} {Cambridge University
  Press},\ \bibinfo {year} {2012})\ \bibinfo {note} {cambridge Books
  Online}\BibitemShut {NoStop}%
\bibitem [{\citenamefont {Lieb}\ and\ \citenamefont {Meixner}(2001)}]{Lieb:01}%
  \BibitemOpen
  \bibfield  {author} {\bibinfo {author} {\bibfnamefont {M.~A.}\ \bibnamefont
  {Lieb}}\ and\ \bibinfo {author} {\bibfnamefont {A.~J.}\ \bibnamefont
  {Meixner}},\ }\href {\doibase 10.1364/OE.8.000458} {\bibfield  {journal}
  {\bibinfo  {journal} {Opt. Express}\ }\textbf {\bibinfo {volume} {8}},\
  \bibinfo {pages} {458} (\bibinfo {year} {2001})}\BibitemShut {NoStop}%
\bibitem [{\citenamefont {Lieb}\ \emph {et~al.}(2004)\citenamefont {Lieb},
  \citenamefont {Zavislan},\ and\ \citenamefont {Novotny}}]{Lieb:04}%
  \BibitemOpen
  \bibfield  {author} {\bibinfo {author} {\bibfnamefont {M.~A.}\ \bibnamefont
  {Lieb}}, \bibinfo {author} {\bibfnamefont {J.~M.}\ \bibnamefont {Zavislan}},
  \ and\ \bibinfo {author} {\bibfnamefont {L.}~\bibnamefont {Novotny}},\ }\href
  {\doibase 10.1364/JOSAB.21.001210} {\bibfield  {journal} {\bibinfo  {journal}
  {J. Opt. Soc. Am. B}\ }\textbf {\bibinfo {volume} {21}},\ \bibinfo {pages}
  {1210} (\bibinfo {year} {2004})}\BibitemShut {NoStop}%
\bibitem [{\citenamefont {Anger}\ \emph {et~al.}(2006)\citenamefont {Anger},
  \citenamefont {Bharadwaj},\ and\ \citenamefont
  {Novotny}}]{PhysRevLett.96.113002}%
  \BibitemOpen
  \bibfield  {author} {\bibinfo {author} {\bibfnamefont {P.}~\bibnamefont
  {Anger}}, \bibinfo {author} {\bibfnamefont {P.}~\bibnamefont {Bharadwaj}}, \
  and\ \bibinfo {author} {\bibfnamefont {L.}~\bibnamefont {Novotny}},\ }\href
  {\doibase 10.1103/PhysRevLett.96.113002} {\bibfield  {journal} {\bibinfo
  {journal} {Phys. Rev. Lett.}\ }\textbf {\bibinfo {volume} {96}},\ \bibinfo
  {pages} {113002} (\bibinfo {year} {2006})}\BibitemShut {NoStop}%
\bibitem [{\citenamefont {Gutbrod}\ \emph {et~al.}(2010)\citenamefont
  {Gutbrod}, \citenamefont {Khoptyar}, \citenamefont {Steiner}, \citenamefont
  {Chizhik}, \citenamefont {Chizhik}, \citenamefont {B{\"a}r},\ and\
  \citenamefont {Meixner}}]{Gutbrod:2010aa}%
  \BibitemOpen
  \bibfield  {author} {\bibinfo {author} {\bibfnamefont {R.}~\bibnamefont
  {Gutbrod}}, \bibinfo {author} {\bibfnamefont {D.}~\bibnamefont {Khoptyar}},
  \bibinfo {author} {\bibfnamefont {M.}~\bibnamefont {Steiner}}, \bibinfo
  {author} {\bibfnamefont {A.~M.}\ \bibnamefont {Chizhik}}, \bibinfo {author}
  {\bibfnamefont {A.~I.}\ \bibnamefont {Chizhik}}, \bibinfo {author}
  {\bibfnamefont {S.}~\bibnamefont {B{\"a}r}}, \ and\ \bibinfo {author}
  {\bibfnamefont {A.~J.}\ \bibnamefont {Meixner}},\ }\bibfield  {booktitle}
  {\emph {\bibinfo {booktitle} {Nano Letters}},\ }\href {\doibase
  10.1021/nl903318p} {\bibfield  {journal} {\bibinfo  {journal} {Nano Letters}\
  }\textbf {\bibinfo {volume} {10}},\ \bibinfo {pages} {504} (\bibinfo {year}
  {2010})}\BibitemShut {NoStop}%
\bibitem [{\citenamefont {Z{\"u}chner}\ \emph {et~al.}(2011)\citenamefont
  {Z{\"u}chner}, \citenamefont {Failla},\ and\ \citenamefont
  {Meixner}}]{Zuchner:2011aa}%
  \BibitemOpen
  \bibfield  {author} {\bibinfo {author} {\bibfnamefont {T.}~\bibnamefont
  {Z{\"u}chner}}, \bibinfo {author} {\bibfnamefont {A.~V.}\ \bibnamefont
  {Failla}}, \ and\ \bibinfo {author} {\bibfnamefont {A.~J.}\ \bibnamefont
  {Meixner}},\ }\bibfield  {booktitle} {\emph {\bibinfo {booktitle} {Angewandte
  Chemie International Edition}},\ }\href {\doibase 10.1002/anie.201005845}
  {\bibfield  {journal} {\bibinfo  {journal} {Angewandte Chemie International
  Edition}\ }\textbf {\bibinfo {volume} {50}},\ \bibinfo {pages} {5274}
  (\bibinfo {year} {2011})}\BibitemShut {NoStop}%
\bibitem [{\citenamefont {Sancho-Parramon}\ and\ \citenamefont
  {Bosch}(2012)}]{Sancho-Parramon:2012aa}%
  \BibitemOpen
  \bibfield  {author} {\bibinfo {author} {\bibfnamefont {J.}~\bibnamefont
  {Sancho-Parramon}}\ and\ \bibinfo {author} {\bibfnamefont {S.}~\bibnamefont
  {Bosch}},\ }\bibfield  {booktitle} {\emph {\bibinfo {booktitle} {ACS Nano}},\
  }\href {\doibase 10.1021/nn303243p} {\bibfield  {journal} {\bibinfo
  {journal} {ACS Nano}\ }\textbf {\bibinfo {volume} {6}},\ \bibinfo {pages}
  {8415} (\bibinfo {year} {2012})}\BibitemShut {NoStop}%
\bibitem [{\citenamefont {Wackenhut}\ \emph {et~al.}(2012)\citenamefont
  {Wackenhut}, \citenamefont {Virgilio~Failla}, \citenamefont {Z{\"u}chner},
  \citenamefont {Steiner},\ and\ \citenamefont {Meixner}}]{Wackenhut:2012aa}%
  \BibitemOpen
  \bibfield  {author} {\bibinfo {author} {\bibfnamefont {F.}~\bibnamefont
  {Wackenhut}}, \bibinfo {author} {\bibfnamefont {A.}~\bibnamefont
  {Virgilio~Failla}}, \bibinfo {author} {\bibfnamefont {T.}~\bibnamefont
  {Z{\"u}chner}}, \bibinfo {author} {\bibfnamefont {M.}~\bibnamefont
  {Steiner}}, \ and\ \bibinfo {author} {\bibfnamefont {A.~J.}\ \bibnamefont
  {Meixner}},\ }\bibfield  {booktitle} {\emph {\bibinfo {booktitle} {Applied
  Physics Letters}},\ }\href {\doibase 10.1063/1.4729152} {\bibfield  {journal}
  {\bibinfo  {journal} {Applied Physics Letters}\ }\textbf {\bibinfo {volume}
  {100}},\ \bibinfo {pages} {263102} (\bibinfo {year} {2012})}\BibitemShut
  {NoStop}%
\bibitem [{\citenamefont {Brewer}\ \emph {et~al.}(2017)\citenamefont {Brewer},
  \citenamefont {Buckholtz}, \citenamefont {Simmons}, \citenamefont {Mueller},\
  and\ \citenamefont {Yavuz}}]{PhysRevX.7.011005}%
  \BibitemOpen
  \bibfield  {author} {\bibinfo {author} {\bibfnamefont {N.~R.}\ \bibnamefont
  {Brewer}}, \bibinfo {author} {\bibfnamefont {Z.~N.}\ \bibnamefont
  {Buckholtz}}, \bibinfo {author} {\bibfnamefont {Z.~J.}\ \bibnamefont
  {Simmons}}, \bibinfo {author} {\bibfnamefont {E.~A.}\ \bibnamefont
  {Mueller}}, \ and\ \bibinfo {author} {\bibfnamefont {D.~D.}\ \bibnamefont
  {Yavuz}},\ }\href {\doibase 10.1103/PhysRevX.7.011005} {\bibfield  {journal}
  {\bibinfo  {journal} {Phys. Rev. X}\ }\textbf {\bibinfo {volume} {7}},\
  \bibinfo {pages} {011005} (\bibinfo {year} {2017})}\BibitemShut {NoStop}%
\bibitem [{\citenamefont {Hasan}\ and\ \citenamefont
  {Kane}(2010)}]{RevModPhys.82.3045}%
  \BibitemOpen
  \bibfield  {author} {\bibinfo {author} {\bibfnamefont {M.~Z.}\ \bibnamefont
  {Hasan}}\ and\ \bibinfo {author} {\bibfnamefont {C.~L.}\ \bibnamefont
  {Kane}},\ }\href {\doibase 10.1103/RevModPhys.82.3045} {\bibfield  {journal}
  {\bibinfo  {journal} {Rev. Mod. Phys.}\ }\textbf {\bibinfo {volume} {82}},\
  \bibinfo {pages} {3045} (\bibinfo {year} {2010})}\BibitemShut {NoStop}%
\bibitem [{\citenamefont {Qi}\ and\ \citenamefont
  {Zhang}(2011)}]{RevModPhys.83.1057}%
  \BibitemOpen
  \bibfield  {author} {\bibinfo {author} {\bibfnamefont {X.-L.}\ \bibnamefont
  {Qi}}\ and\ \bibinfo {author} {\bibfnamefont {S.-C.}\ \bibnamefont {Zhang}},\
  }\href {\doibase 10.1103/RevModPhys.83.1057} {\bibfield  {journal} {\bibinfo
  {journal} {Rev. Mod. Phys.}\ }\textbf {\bibinfo {volume} {83}},\ \bibinfo
  {pages} {1057} (\bibinfo {year} {2011})}\BibitemShut {NoStop}%
\bibitem [{\citenamefont {Bukov}\ \emph {et~al.}(2015)\citenamefont {Bukov},
  \citenamefont {D'Alessio},\ and\ \citenamefont
  {Polkovnikov}}]{doi:10.1080/00018732.2015.1055918}%
  \BibitemOpen
  \bibfield  {author} {\bibinfo {author} {\bibfnamefont {M.}~\bibnamefont
  {Bukov}}, \bibinfo {author} {\bibfnamefont {L.}~\bibnamefont {D'Alessio}}, \
  and\ \bibinfo {author} {\bibfnamefont {A.}~\bibnamefont {Polkovnikov}},\
  }\href {\doibase 10.1080/00018732.2015.1055918} {\bibfield  {journal}
  {\bibinfo  {journal} {Advances in Physics}\ }\textbf {\bibinfo {volume}
  {64}},\ \bibinfo {pages} {139} (\bibinfo {year} {2015})}\BibitemShut
  {NoStop}%
\bibitem [{\citenamefont {Eckardt}(2017)}]{RevModPhys.89.011004}%
  \BibitemOpen
  \bibfield  {author} {\bibinfo {author} {\bibfnamefont {A.}~\bibnamefont
  {Eckardt}},\ }\href {\doibase 10.1103/RevModPhys.89.011004} {\bibfield
  {journal} {\bibinfo  {journal} {Rev. Mod. Phys.}\ }\textbf {\bibinfo {volume}
  {89}},\ \bibinfo {pages} {011004} (\bibinfo {year} {2017})}\BibitemShut
  {NoStop}%
\bibitem [{\citenamefont {Oka}\ and\ \citenamefont
  {Kitamura}(2018)}]{kitamura-oka}%
  \BibitemOpen
  \bibfield  {author} {\bibinfo {author} {\bibfnamefont {T.}~\bibnamefont
  {Oka}}\ and\ \bibinfo {author} {\bibfnamefont {S.}~\bibnamefont {Kitamura}},\
  }\href@noop {} {\bibfield  {journal} {\bibinfo  {journal} {arXiv:1804.03212}\
  } (\bibinfo {year} {2018})}\BibitemShut {NoStop}%
\bibitem [{\citenamefont {Sato}\ \emph {et~al.}(2016)\citenamefont {Sato},
  \citenamefont {Takayoshi},\ and\ \citenamefont {Oka}}]{arXiv:1602.03702}%
  \BibitemOpen
  \bibfield  {author} {\bibinfo {author} {\bibfnamefont {M.}~\bibnamefont
  {Sato}}, \bibinfo {author} {\bibfnamefont {S.}~\bibnamefont {Takayoshi}}, \
  and\ \bibinfo {author} {\bibfnamefont {T.}~\bibnamefont {Oka}},\ }\href@noop
  {} {\bibfield  {journal} {\bibinfo  {journal} {Phys. Rev. Lett.}\ }\textbf
  {\bibinfo {volume} {117}},\ \bibinfo {pages} {147202} (\bibinfo {year}
  {2016})}\BibitemShut {NoStop}%
\bibitem [{\citenamefont {Higashikawa}\ \emph {et~al.}(2018)\citenamefont
  {Higashikawa}, \citenamefont {Fujita},\ and\ \citenamefont
  {Sato}}]{higashikawa}%
  \BibitemOpen
  \bibfield  {author} {\bibinfo {author} {\bibfnamefont {S.}~\bibnamefont
  {Higashikawa}}, \bibinfo {author} {\bibfnamefont {H.}~\bibnamefont {Fujita}},
  \ and\ \bibinfo {author} {\bibfnamefont {M.}~\bibnamefont {Sato}},\
  }\href@noop {} {\bibfield  {journal} {\bibinfo  {journal} {arXiv:1810.01103}\
  } (\bibinfo {year} {2018})}\BibitemShut {NoStop}%
\bibitem [{\citenamefont {Rittweger}\ \emph {et~al.}(2009)\citenamefont
  {Rittweger}, \citenamefont {Han}, \citenamefont {Irvine}, \citenamefont
  {Eggeling},\ and\ \citenamefont {Hell}}]{Rittweger:2009aa}%
  \BibitemOpen
  \bibfield  {author} {\bibinfo {author} {\bibfnamefont {E.}~\bibnamefont
  {Rittweger}}, \bibinfo {author} {\bibfnamefont {K.~Y.}\ \bibnamefont {Han}},
  \bibinfo {author} {\bibfnamefont {S.~E.}\ \bibnamefont {Irvine}}, \bibinfo
  {author} {\bibfnamefont {C.}~\bibnamefont {Eggeling}}, \ and\ \bibinfo
  {author} {\bibfnamefont {S.~W.}\ \bibnamefont {Hell}},\ }\href
  {http://dx.doi.org/10.1038/nphoton.2009.2} {\bibfield  {journal} {\bibinfo
  {journal} {Nature Photonics}\ }\textbf {\bibinfo {volume} {3}},\ \bibinfo
  {pages} {144 EP } (\bibinfo {year} {2009})}\BibitemShut {NoStop}%
\bibitem [{\citenamefont {Hamazaki}\ \emph {et~al.}(2010)\citenamefont
  {Hamazaki}, \citenamefont {Morita}, \citenamefont {Chujo}, \citenamefont
  {Kobayashi}, \citenamefont {Tanda},\ and\ \citenamefont
  {Omatsu}}]{Hamazaki:10}%
  \BibitemOpen
  \bibfield  {author} {\bibinfo {author} {\bibfnamefont {J.}~\bibnamefont
  {Hamazaki}}, \bibinfo {author} {\bibfnamefont {R.}~\bibnamefont {Morita}},
  \bibinfo {author} {\bibfnamefont {K.}~\bibnamefont {Chujo}}, \bibinfo
  {author} {\bibfnamefont {Y.}~\bibnamefont {Kobayashi}}, \bibinfo {author}
  {\bibfnamefont {S.}~\bibnamefont {Tanda}}, \ and\ \bibinfo {author}
  {\bibfnamefont {T.}~\bibnamefont {Omatsu}},\ }\href {\doibase
  10.1364/OE.18.002144} {\bibfield  {journal} {\bibinfo  {journal} {Opt.
  Express}\ }\textbf {\bibinfo {volume} {18}},\ \bibinfo {pages} {2144}
  (\bibinfo {year} {2010})}\BibitemShut {NoStop}%
\bibitem [{\citenamefont {Terhalle}\ \emph {et~al.}(2011)\citenamefont
  {Terhalle}, \citenamefont {Langner}, \citenamefont {P\"{a}iv\"{a}nranta},
  \citenamefont {Guzenko}, \citenamefont {David},\ and\ \citenamefont
  {Ekinci}}]{Terhalle:11}%
  \BibitemOpen
  \bibfield  {author} {\bibinfo {author} {\bibfnamefont {B.}~\bibnamefont
  {Terhalle}}, \bibinfo {author} {\bibfnamefont {A.}~\bibnamefont {Langner}},
  \bibinfo {author} {\bibfnamefont {B.}~\bibnamefont {P\"{a}iv\"{a}nranta}},
  \bibinfo {author} {\bibfnamefont {V.~A.}\ \bibnamefont {Guzenko}}, \bibinfo
  {author} {\bibfnamefont {C.}~\bibnamefont {David}}, \ and\ \bibinfo {author}
  {\bibfnamefont {Y.}~\bibnamefont {Ekinci}},\ }\href {\doibase
  10.1364/OL.36.004143} {\bibfield  {journal} {\bibinfo  {journal} {Opt.
  Lett.}\ }\textbf {\bibinfo {volume} {36}},\ \bibinfo {pages} {4143} (\bibinfo
  {year} {2011})}\BibitemShut {NoStop}%
\bibitem [{\citenamefont {Toyoda}\ \emph {et~al.}(2012)\citenamefont {Toyoda},
  \citenamefont {Miyamoto}, \citenamefont {Aoki}, \citenamefont {Morita},\ and\
  \citenamefont {Omatsu}}]{Toyoda:2012aa}%
  \BibitemOpen
  \bibfield  {author} {\bibinfo {author} {\bibfnamefont {K.}~\bibnamefont
  {Toyoda}}, \bibinfo {author} {\bibfnamefont {K.}~\bibnamefont {Miyamoto}},
  \bibinfo {author} {\bibfnamefont {N.}~\bibnamefont {Aoki}}, \bibinfo {author}
  {\bibfnamefont {R.}~\bibnamefont {Morita}}, \ and\ \bibinfo {author}
  {\bibfnamefont {T.}~\bibnamefont {Omatsu}},\ }\href {\doibase
  10.1021/nl301347j} {\bibfield  {journal} {\bibinfo  {journal} {Nano Lett.}\
  }\textbf {\bibinfo {volume} {12}},\ \bibinfo {pages} {3645} (\bibinfo {year}
  {2012})}\BibitemShut {NoStop}%
\bibitem [{\citenamefont {Takahashi}\ \emph {et~al.}(2013)\citenamefont
  {Takahashi}, \citenamefont {Toyoda}, \citenamefont {Takizawa}, \citenamefont
  {Miyamoto}, \citenamefont {Morita},\ and\ \citenamefont
  {Omatsu}}]{Takahashi:13}%
  \BibitemOpen
  \bibfield  {author} {\bibinfo {author} {\bibfnamefont {F.}~\bibnamefont
  {Takahashi}}, \bibinfo {author} {\bibfnamefont {K.}~\bibnamefont {Toyoda}},
  \bibinfo {author} {\bibfnamefont {S.}~\bibnamefont {Takizawa}}, \bibinfo
  {author} {\bibfnamefont {K.}~\bibnamefont {Miyamoto}}, \bibinfo {author}
  {\bibfnamefont {R.}~\bibnamefont {Morita}}, \ and\ \bibinfo {author}
  {\bibfnamefont {T.}~\bibnamefont {Omatsu}},\ }in\ \href {\doibase
  10.1364/CLEO_SI.2013.CM3H.6} {\emph {\bibinfo {booktitle} {CLEO: 2013}}}\
  (\bibinfo  {publisher} {Optical Society of America},\ \bibinfo {year}
  {2013})\ p.\ \bibinfo {pages} {CM3H.6}\BibitemShut {NoStop}%
\bibitem [{\citenamefont {Shigematsu}\ \emph {et~al.}(2016)\citenamefont
  {Shigematsu}, \citenamefont {Yamane}, \citenamefont {Morita},\ and\
  \citenamefont {Toda}}]{PhysRevB.93.045205}%
  \BibitemOpen
  \bibfield  {author} {\bibinfo {author} {\bibfnamefont {K.}~\bibnamefont
  {Shigematsu}}, \bibinfo {author} {\bibfnamefont {K.}~\bibnamefont {Yamane}},
  \bibinfo {author} {\bibfnamefont {R.}~\bibnamefont {Morita}}, \ and\ \bibinfo
  {author} {\bibfnamefont {Y.}~\bibnamefont {Toda}},\ }\href {\doibase
  10.1103/PhysRevB.93.045205} {\bibfield  {journal} {\bibinfo  {journal} {Phys.
  Rev. B}\ }\textbf {\bibinfo {volume} {93}},\ \bibinfo {pages} {045205}
  (\bibinfo {year} {2016})}\BibitemShut {NoStop}%
\bibitem [{\citenamefont {Fujita}\ and\ \citenamefont
  {Sato}(2017{\natexlab{a}})}]{Fujita2016}%
  \BibitemOpen
  \bibfield  {author} {\bibinfo {author} {\bibfnamefont {H.}~\bibnamefont
  {Fujita}}\ and\ \bibinfo {author} {\bibfnamefont {M.}~\bibnamefont {Sato}},\
  }\href {\doibase 10.1103/PhysRevB.95.054421} {\bibfield  {journal} {\bibinfo
  {journal} {Phys. Rev. B}\ }\textbf {\bibinfo {volume} {95}},\ \bibinfo
  {pages} {054421} (\bibinfo {year} {2017}{\natexlab{a}})}\BibitemShut
  {NoStop}%
\bibitem [{\citenamefont {Fujita}\ and\ \citenamefont
  {Sato}(2017{\natexlab{b}})}]{PhysRevB.96.060407}%
  \BibitemOpen
  \bibfield  {author} {\bibinfo {author} {\bibfnamefont {H.}~\bibnamefont
  {Fujita}}\ and\ \bibinfo {author} {\bibfnamefont {M.}~\bibnamefont {Sato}},\
  }\href {\doibase 10.1103/PhysRevB.96.060407} {\bibfield  {journal} {\bibinfo
  {journal} {Phys. Rev. B}\ }\textbf {\bibinfo {volume} {96}},\ \bibinfo
  {pages} {060407} (\bibinfo {year} {2017}{\natexlab{b}})}\BibitemShut
  {NoStop}%
\bibitem [{Note1()}]{Note1}%
  \BibitemOpen
  \bibinfo {note} {Very recently, two of the authors proposed its application
  for the Fermi surface measurement of magnetic metals and generic metals under
  ultra-high pressure~\cite {Fujita:2018aa}}\BibitemShut {NoStop}%
\bibitem [{Note2()}]{Note2}%
  \BibitemOpen
  \bibinfo {note} {The electromagnetic field configuration of focused CVBs is
  qualitatively the same as the TE011 cavity mode of microwave~\cite
  {Hertel:2005aa} which is already the standard of ESR at microwave
  frequencies. Focused CVBs are, therefore, providing an optical way of
  mimicking the electromagnetic configuration of TE011 mode.}\BibitemShut
  {Stop}%
\bibitem [{\citenamefont {Wertz}(2012)}]{wertz2012electron}%
  \BibitemOpen
  \bibfield  {author} {\bibinfo {author} {\bibfnamefont {J.}~\bibnamefont
  {Wertz}},\ }\href {https://books.google.co.jp/books?id=myXpCAAAQBAJ} {\emph
  {\bibinfo {title} {Electron Spin Resonance: Elementary Theory and Practical
  Applications}}}\ (\bibinfo  {publisher} {Springer Netherlands},\ \bibinfo
  {year} {2012})\BibitemShut {NoStop}%
\bibitem [{\citenamefont {Slichter}(1996)}]{slichter1996principles}%
  \BibitemOpen
  \bibfield  {author} {\bibinfo {author} {\bibfnamefont {C.}~\bibnamefont
  {Slichter}},\ }\href {https://books.google.co.jp/books?id=zgnrRkaIhFoC}
  {\emph {\bibinfo {title} {Principles of Magnetic Resonance}}},\ Springer
  Series in Solid-State Sciences\ (\bibinfo  {publisher} {Springer Berlin
  Heidelberg},\ \bibinfo {year} {1996})\BibitemShut {NoStop}%
\bibitem [{\citenamefont {Ashcroft}\ and\ \citenamefont
  {Mermin}(1976)}]{Ashcroft}%
  \BibitemOpen
  \bibfield  {author} {\bibinfo {author} {\bibfnamefont {N.}~\bibnamefont
  {Ashcroft}}\ and\ \bibinfo {author} {\bibfnamefont {N.}~\bibnamefont
  {Mermin}},\ }\href@noop {} {\emph {\bibinfo {title} {{Solid State
  Physics}}}}\ (\bibinfo  {publisher} {Saunders College},\ \bibinfo {address}
  {Philadelphia},\ \bibinfo {year} {1976})\BibitemShut {NoStop}%
\bibitem [{\citenamefont {Katsumata}(2000)}]{0953-8984-12-47-201}%
  \BibitemOpen
  \bibfield  {author} {\bibinfo {author} {\bibfnamefont {K.}~\bibnamefont
  {Katsumata}},\ }\href {http://stacks.iop.org/0953-8984/12/i=47/a=201}
  {\bibfield  {journal} {\bibinfo  {journal} {Journal of Physics: Condensed
  Matter}\ }\textbf {\bibinfo {volume} {12}},\ \bibinfo {pages} {R589}
  (\bibinfo {year} {2000})}\BibitemShut {NoStop}%
\bibitem [{\citenamefont {Bennati}\ and\ \citenamefont
  {Prisner}(2005)}]{0034-4885-68-2-R05}%
  \BibitemOpen
  \bibfield  {author} {\bibinfo {author} {\bibfnamefont {M.}~\bibnamefont
  {Bennati}}\ and\ \bibinfo {author} {\bibfnamefont {T.~F.}\ \bibnamefont
  {Prisner}},\ }\href {http://stacks.iop.org/0034-4885/68/i=2/a=R05} {\bibfield
   {journal} {\bibinfo  {journal} {Reports on Progress in Physics}\ }\textbf
  {\bibinfo {volume} {68}},\ \bibinfo {pages} {411} (\bibinfo {year}
  {2005})}\BibitemShut {NoStop}%
\bibitem [{\citenamefont {Grinberg}\ and\ \citenamefont
  {Berliner}(2013)}]{grinberg2013very}%
  \BibitemOpen
  \bibfield  {author} {\bibinfo {author} {\bibfnamefont {O.}~\bibnamefont
  {Grinberg}}\ and\ \bibinfo {author} {\bibfnamefont {L.}~\bibnamefont
  {Berliner}},\ }\href {https://books.google.co.jp/books?id=N7vVBwAAQBAJ}
  {\emph {\bibinfo {title} {{Very High Frequency (VHF) ESR/EPR}}}},\ Biological
  Magnetic Resonance\ (\bibinfo  {publisher} {Springer US},\ \bibinfo {year}
  {2013})\BibitemShut {NoStop}%
\bibitem [{\citenamefont {Kampfrath}\ \emph {et~al.}(2011)\citenamefont
  {Kampfrath}, \citenamefont {Sell}, \citenamefont {Klatt}, \citenamefont
  {Pashkin}, \citenamefont {Mahrlein}, \citenamefont {Dekorsy}, \citenamefont
  {Wolf}, \citenamefont {Fiebig}, \citenamefont {Leitenstorfer},\ and\
  \citenamefont {Huber}}]{Kampfrath:2011aa}%
  \BibitemOpen
  \bibfield  {author} {\bibinfo {author} {\bibfnamefont {T.}~\bibnamefont
  {Kampfrath}}, \bibinfo {author} {\bibfnamefont {A.}~\bibnamefont {Sell}},
  \bibinfo {author} {\bibfnamefont {G.}~\bibnamefont {Klatt}}, \bibinfo
  {author} {\bibfnamefont {A.}~\bibnamefont {Pashkin}}, \bibinfo {author}
  {\bibfnamefont {S.}~\bibnamefont {Mahrlein}}, \bibinfo {author}
  {\bibfnamefont {T.}~\bibnamefont {Dekorsy}}, \bibinfo {author} {\bibfnamefont
  {M.}~\bibnamefont {Wolf}}, \bibinfo {author} {\bibfnamefont {M.}~\bibnamefont
  {Fiebig}}, \bibinfo {author} {\bibfnamefont {A.}~\bibnamefont
  {Leitenstorfer}}, \ and\ \bibinfo {author} {\bibfnamefont {R.}~\bibnamefont
  {Huber}},\ }\href {http://dx.doi.org/10.1038/nphoton.2010.259} {\bibfield
  {journal} {\bibinfo  {journal} {Nat. Photon.}\ }\textbf {\bibinfo {volume}
  {5}},\ \bibinfo {pages} {31} (\bibinfo {year} {2011})}\BibitemShut {NoStop}%
\bibitem [{\citenamefont {Jin}\ \emph {et~al.}(2013)\citenamefont {Jin},
  \citenamefont {Mics}, \citenamefont {Ma}, \citenamefont {Cheng},
  \citenamefont {Bonn},\ and\ \citenamefont
  {Turchinovich}}]{PhysRevB.87.094422}%
  \BibitemOpen
  \bibfield  {author} {\bibinfo {author} {\bibfnamefont {Z.}~\bibnamefont
  {Jin}}, \bibinfo {author} {\bibfnamefont {Z.}~\bibnamefont {Mics}}, \bibinfo
  {author} {\bibfnamefont {G.}~\bibnamefont {Ma}}, \bibinfo {author}
  {\bibfnamefont {Z.}~\bibnamefont {Cheng}}, \bibinfo {author} {\bibfnamefont
  {M.}~\bibnamefont {Bonn}}, \ and\ \bibinfo {author} {\bibfnamefont
  {D.}~\bibnamefont {Turchinovich}},\ }\href {\doibase
  10.1103/PhysRevB.87.094422} {\bibfield  {journal} {\bibinfo  {journal} {Phys.
  Rev. B}\ }\textbf {\bibinfo {volume} {87}},\ \bibinfo {pages} {094422}
  (\bibinfo {year} {2013})}\BibitemShut {NoStop}%
\bibitem [{\citenamefont {Savary}\ and\ \citenamefont
  {Balents}(2017)}]{0034-4885-80-1-016502}%
  \BibitemOpen
  \bibfield  {author} {\bibinfo {author} {\bibfnamefont {L.}~\bibnamefont
  {Savary}}\ and\ \bibinfo {author} {\bibfnamefont {L.}~\bibnamefont
  {Balents}},\ }\href {http://stacks.iop.org/0034-4885/80/i=1/a=016502}
  {\bibfield  {journal} {\bibinfo  {journal} {Reports on Progress in Physics}\
  }\textbf {\bibinfo {volume} {80}},\ \bibinfo {pages} {016502} (\bibinfo
  {year} {2017})}\BibitemShut {NoStop}%
\bibitem [{\citenamefont {Rashba}(2008)}]{PhysRevB.78.195302}%
  \BibitemOpen
  \bibfield  {author} {\bibinfo {author} {\bibfnamefont {E.~I.}\ \bibnamefont
  {Rashba}},\ }\href {\doibase 10.1103/PhysRevB.78.195302} {\bibfield
  {journal} {\bibinfo  {journal} {Phys. Rev. B}\ }\textbf {\bibinfo {volume}
  {78}},\ \bibinfo {pages} {195302} (\bibinfo {year} {2008})}\BibitemShut
  {NoStop}%
\bibitem [{\citenamefont {Nowack}\ \emph {et~al.}(2007)\citenamefont {Nowack},
  \citenamefont {Koppens}, \citenamefont {Nazarov},\ and\ \citenamefont
  {Vandersypen}}]{Nowack1430}%
  \BibitemOpen
  \bibfield  {author} {\bibinfo {author} {\bibfnamefont {K.~C.}\ \bibnamefont
  {Nowack}}, \bibinfo {author} {\bibfnamefont {F.~H.~L.}\ \bibnamefont
  {Koppens}}, \bibinfo {author} {\bibfnamefont {Y.~V.}\ \bibnamefont
  {Nazarov}}, \ and\ \bibinfo {author} {\bibfnamefont {L.~M.~K.}\ \bibnamefont
  {Vandersypen}},\ }\href {\doibase 10.1126/science.1148092} {\bibfield
  {journal} {\bibinfo  {journal} {Science}\ }\textbf {\bibinfo {volume}
  {318}},\ \bibinfo {pages} {1430} (\bibinfo {year} {2007})}\BibitemShut
  {NoStop}%
\bibitem [{\citenamefont {Bolens}\ \emph {et~al.}(2017)\citenamefont {Bolens},
  \citenamefont {Katsura}, \citenamefont {Ogata},\ and\ \citenamefont
  {Miyashita}}]{PhysRevB.95.235115}%
  \BibitemOpen
  \bibfield  {author} {\bibinfo {author} {\bibfnamefont {A.}~\bibnamefont
  {Bolens}}, \bibinfo {author} {\bibfnamefont {H.}~\bibnamefont {Katsura}},
  \bibinfo {author} {\bibfnamefont {M.}~\bibnamefont {Ogata}}, \ and\ \bibinfo
  {author} {\bibfnamefont {S.}~\bibnamefont {Miyashita}},\ }\href {\doibase
  10.1103/PhysRevB.95.235115} {\bibfield  {journal} {\bibinfo  {journal} {Phys.
  Rev. B}\ }\textbf {\bibinfo {volume} {95}},\ \bibinfo {pages} {235115}
  (\bibinfo {year} {2017})}\BibitemShut {NoStop}%
\bibitem [{Note3()}]{Note3}%
  \BibitemOpen
  \bibinfo {note} {Here we omit possible cross-terms of $\protect \bm {E}$ and
  $\protect \bm {B}$. For example, in multiferroic materials, they could be
  important.}\BibitemShut {Stop}%
\bibitem [{Note4()}]{Note4}%
  \BibitemOpen
  \bibinfo {note} {Since the wavelength of a light at 1 THz is much longer than
  the lattice constant, field absorption is locally the same for Gaussian beams
  and CVB. Their difference comes only from their field
  distributions.}\BibitemShut {Stop}%
\bibitem [{Note5()}]{Note5}%
  \BibitemOpen
  \bibinfo {note} {The Drude model of the electronic conduction predicts
  $\sigma '(\omega ) = \sigma _0/(1 + \omega ^2 \tau ^2)$. The relaxation time
  of conduction electrons $\tau $ is typically order of femtosecond and thus
  the product $\omega \tau $ is, for $\omega $ in the THz region, quite
  small.}\BibitemShut {Stop}%
\bibitem [{\citenamefont {Torrey}(1956)}]{PhysRev.104.563}%
  \BibitemOpen
  \bibfield  {author} {\bibinfo {author} {\bibfnamefont {H.~C.}\ \bibnamefont
  {Torrey}},\ }\href {\doibase 10.1103/PhysRev.104.563} {\bibfield  {journal}
  {\bibinfo  {journal} {Phys. Rev.}\ }\textbf {\bibinfo {volume} {104}},\
  \bibinfo {pages} {563} (\bibinfo {year} {1956})}\BibitemShut {NoStop}%
\bibitem [{\citenamefont {Semiconductor.}(2018)}]{semicon_britannica}%
  \BibitemOpen
  \bibfield  {author} {\bibinfo {author} {\bibnamefont {Semiconductor.}},\
  }\href@noop {} {\emph {\bibinfo {title} {{Encyclopedia Britannica}}}}\
  (\bibinfo  {publisher} {Encyclopedia Britannica, inc.},\ \bibinfo {year}
  {2018})\BibitemShut {NoStop}%
\bibitem [{\citenamefont {Barnes}\ and\ \citenamefont
  {Freed}(1997)}]{Barnes:1997aa}%
  \BibitemOpen
  \bibfield  {author} {\bibinfo {author} {\bibfnamefont {J.~P.}\ \bibnamefont
  {Barnes}}\ and\ \bibinfo {author} {\bibfnamefont {J.~H.}\ \bibnamefont
  {Freed}},\ }\bibfield  {booktitle} {\emph {\bibinfo {booktitle} {Review of
  Scientific Instruments}},\ }\href {\doibase 10.1063/1.1148205} {\bibfield
  {journal} {\bibinfo  {journal} {Review of Scientific Instruments}\ }\textbf
  {\bibinfo {volume} {68}},\ \bibinfo {pages} {2838} (\bibinfo {year}
  {1997})}\BibitemShut {NoStop}%
\bibitem [{\citenamefont {Reijerse}(2009)}]{Reijerse2009}%
  \BibitemOpen
  \bibfield  {author} {\bibinfo {author} {\bibfnamefont {E.~J.}\ \bibnamefont
  {Reijerse}},\ }\href {\doibase 10.1007/s00723-009-0070-y} {\bibfield
  {journal} {\bibinfo  {journal} {Applied Magnetic Resonance}\ }\textbf
  {\bibinfo {volume} {37}},\ \bibinfo {pages} {795} (\bibinfo {year}
  {2009})}\BibitemShut {NoStop}%
\bibitem [{\citenamefont {Sichelschmidt}\ \emph {et~al.}(2003)\citenamefont
  {Sichelschmidt}, \citenamefont {Ivanshin}, \citenamefont {Ferstl},
  \citenamefont {Geibel},\ and\ \citenamefont
  {Steglich}}]{PhysRevLett.91.156401}%
  \BibitemOpen
  \bibfield  {author} {\bibinfo {author} {\bibfnamefont {J.}~\bibnamefont
  {Sichelschmidt}}, \bibinfo {author} {\bibfnamefont {V.~A.}\ \bibnamefont
  {Ivanshin}}, \bibinfo {author} {\bibfnamefont {J.}~\bibnamefont {Ferstl}},
  \bibinfo {author} {\bibfnamefont {C.}~\bibnamefont {Geibel}}, \ and\ \bibinfo
  {author} {\bibfnamefont {F.}~\bibnamefont {Steglich}},\ }\href {\doibase
  10.1103/PhysRevLett.91.156401} {\bibfield  {journal} {\bibinfo  {journal}
  {Phys. Rev. Lett.}\ }\textbf {\bibinfo {volume} {91}},\ \bibinfo {pages}
  {156401} (\bibinfo {year} {2003})}\BibitemShut {NoStop}%
\bibitem [{\citenamefont {Schlottmann}(2009)}]{PhysRevB.79.045104}%
  \BibitemOpen
  \bibfield  {author} {\bibinfo {author} {\bibfnamefont {P.}~\bibnamefont
  {Schlottmann}},\ }\href {\doibase 10.1103/PhysRevB.79.045104} {\bibfield
  {journal} {\bibinfo  {journal} {Phys. Rev. B}\ }\textbf {\bibinfo {volume}
  {79}},\ \bibinfo {pages} {045104} (\bibinfo {year} {2009})}\BibitemShut
  {NoStop}%
\bibitem [{\citenamefont {Sinova}\ \emph {et~al.}(2015)\citenamefont {Sinova},
  \citenamefont {Valenzuela}, \citenamefont {Wunderlich}, \citenamefont
  {Back},\ and\ \citenamefont {Jungwirth}}]{RevModPhys.87.1213}%
  \BibitemOpen
  \bibfield  {author} {\bibinfo {author} {\bibfnamefont {J.}~\bibnamefont
  {Sinova}}, \bibinfo {author} {\bibfnamefont {S.~O.}\ \bibnamefont
  {Valenzuela}}, \bibinfo {author} {\bibfnamefont {J.}~\bibnamefont
  {Wunderlich}}, \bibinfo {author} {\bibfnamefont {C.~H.}\ \bibnamefont
  {Back}}, \ and\ \bibinfo {author} {\bibfnamefont {T.}~\bibnamefont
  {Jungwirth}},\ }\href {\doibase 10.1103/RevModPhys.87.1213} {\bibfield
  {journal} {\bibinfo  {journal} {Rev. Mod. Phys.}\ }\textbf {\bibinfo {volume}
  {87}},\ \bibinfo {pages} {1213} (\bibinfo {year} {2015})}\BibitemShut
  {NoStop}%
\bibitem [{\citenamefont {Tokura}\ \emph {et~al.}(2014)\citenamefont {Tokura},
  \citenamefont {Seki},\ and\ \citenamefont {Nagaosa}}]{0034-4885-77-7-076501}%
  \BibitemOpen
  \bibfield  {author} {\bibinfo {author} {\bibfnamefont {Y.}~\bibnamefont
  {Tokura}}, \bibinfo {author} {\bibfnamefont {S.}~\bibnamefont {Seki}}, \ and\
  \bibinfo {author} {\bibfnamefont {N.}~\bibnamefont {Nagaosa}},\ }\href
  {http://stacks.iop.org/0034-4885/77/i=7/a=076501} {\bibfield  {journal}
  {\bibinfo  {journal} {Rep. Prog. Phys.}\ }\textbf {\bibinfo {volume} {77}},\
  \bibinfo {pages} {076501} (\bibinfo {year} {2014})}\BibitemShut {NoStop}%
\bibitem [{\citenamefont {Katsura}\ \emph {et~al.}(2005)\citenamefont
  {Katsura}, \citenamefont {Nagaosa},\ and\ \citenamefont
  {Balatsky}}]{PhysRevLett.95.057205}%
  \BibitemOpen
  \bibfield  {author} {\bibinfo {author} {\bibfnamefont {H.}~\bibnamefont
  {Katsura}}, \bibinfo {author} {\bibfnamefont {N.}~\bibnamefont {Nagaosa}}, \
  and\ \bibinfo {author} {\bibfnamefont {A.~V.}\ \bibnamefont {Balatsky}},\
  }\href {\doibase 10.1103/PhysRevLett.95.057205} {\bibfield  {journal}
  {\bibinfo  {journal} {Phys. Rev. Lett.}\ }\textbf {\bibinfo {volume} {95}},\
  \bibinfo {pages} {057205} (\bibinfo {year} {2005})}\BibitemShut {NoStop}%
\bibitem [{\citenamefont {Mochizuki}\ \emph {et~al.}(2010)\citenamefont
  {Mochizuki}, \citenamefont {Furukawa},\ and\ \citenamefont
  {Nagaosa}}]{PhysRevLett.104.177206}%
  \BibitemOpen
  \bibfield  {author} {\bibinfo {author} {\bibfnamefont {M.}~\bibnamefont
  {Mochizuki}}, \bibinfo {author} {\bibfnamefont {N.}~\bibnamefont {Furukawa}},
  \ and\ \bibinfo {author} {\bibfnamefont {N.}~\bibnamefont {Nagaosa}},\ }\href
  {\doibase 10.1103/PhysRevLett.104.177206} {\bibfield  {journal} {\bibinfo
  {journal} {Phys. Rev. Lett.}\ }\textbf {\bibinfo {volume} {104}},\ \bibinfo
  {pages} {177206} (\bibinfo {year} {2010})}\BibitemShut {NoStop}%
\bibitem [{\citenamefont {Pimenov}\ \emph {et~al.}(2006)\citenamefont
  {Pimenov}, \citenamefont {Mukhin}, \citenamefont {Ivanov}, \citenamefont
  {Travkin}, \citenamefont {Balbashov},\ and\ \citenamefont
  {Loidl}}]{Pimenov:2006aa}%
  \BibitemOpen
  \bibfield  {author} {\bibinfo {author} {\bibfnamefont {A.}~\bibnamefont
  {Pimenov}}, \bibinfo {author} {\bibfnamefont {A.~A.}\ \bibnamefont {Mukhin}},
  \bibinfo {author} {\bibfnamefont {V.~Y.}\ \bibnamefont {Ivanov}}, \bibinfo
  {author} {\bibfnamefont {V.~D.}\ \bibnamefont {Travkin}}, \bibinfo {author}
  {\bibfnamefont {A.~M.}\ \bibnamefont {Balbashov}}, \ and\ \bibinfo {author}
  {\bibfnamefont {A.}~\bibnamefont {Loidl}},\ }\href
  {http://dx.doi.org/10.1038/nphys212} {\bibfield  {journal} {\bibinfo
  {journal} {Nature Phys.}\ }\textbf {\bibinfo {volume} {2}},\ \bibinfo {pages}
  {97} (\bibinfo {year} {2006})}\BibitemShut {NoStop}%
\bibitem [{\citenamefont {Katsura}\ \emph {et~al.}(2007)\citenamefont
  {Katsura}, \citenamefont {Balatsky},\ and\ \citenamefont
  {Nagaosa}}]{PhysRevLett.98.027203}%
  \BibitemOpen
  \bibfield  {author} {\bibinfo {author} {\bibfnamefont {H.}~\bibnamefont
  {Katsura}}, \bibinfo {author} {\bibfnamefont {A.~V.}\ \bibnamefont
  {Balatsky}}, \ and\ \bibinfo {author} {\bibfnamefont {N.}~\bibnamefont
  {Nagaosa}},\ }\href {\doibase 10.1103/PhysRevLett.98.027203} {\bibfield
  {journal} {\bibinfo  {journal} {Phys. Rev. Lett.}\ }\textbf {\bibinfo
  {volume} {98}},\ \bibinfo {pages} {027203} (\bibinfo {year}
  {2007})}\BibitemShut {NoStop}%
\bibitem [{\citenamefont {Takahashi}\ \emph {et~al.}(2012)\citenamefont
  {Takahashi}, \citenamefont {Shimano}, \citenamefont {Kaneko}, \citenamefont
  {Murakawa},\ and\ \citenamefont {Tokura}}]{Takahashi:2012aa}%
  \BibitemOpen
  \bibfield  {author} {\bibinfo {author} {\bibfnamefont {Y.}~\bibnamefont
  {Takahashi}}, \bibinfo {author} {\bibfnamefont {R.}~\bibnamefont {Shimano}},
  \bibinfo {author} {\bibfnamefont {Y.}~\bibnamefont {Kaneko}}, \bibinfo
  {author} {\bibfnamefont {H.}~\bibnamefont {Murakawa}}, \ and\ \bibinfo
  {author} {\bibfnamefont {Y.}~\bibnamefont {Tokura}},\ }\href
  {http://dx.doi.org/10.1038/nphys2161} {\bibfield  {journal} {\bibinfo
  {journal} {Nature Phys.}\ }\textbf {\bibinfo {volume} {8}},\ \bibinfo {pages}
  {121} (\bibinfo {year} {2012})}\BibitemShut {NoStop}%
\bibitem [{\citenamefont {H\"uvonen}\ \emph {et~al.}(2009)\citenamefont
  {H\"uvonen}, \citenamefont {Nagel}, \citenamefont {R\~o\ om}, \citenamefont
  {Choi}, \citenamefont {Zhang}, \citenamefont {Park},\ and\ \citenamefont
  {Cheong}}]{PhysRevB.80.100402}%
  \BibitemOpen
  \bibfield  {author} {\bibinfo {author} {\bibfnamefont {D.}~\bibnamefont
  {H\"uvonen}}, \bibinfo {author} {\bibfnamefont {U.}~\bibnamefont {Nagel}},
  \bibinfo {author} {\bibfnamefont {T.}~\bibnamefont {R\~o\ om}}, \bibinfo
  {author} {\bibfnamefont {Y.~J.}\ \bibnamefont {Choi}}, \bibinfo {author}
  {\bibfnamefont {C.~L.}\ \bibnamefont {Zhang}}, \bibinfo {author}
  {\bibfnamefont {S.}~\bibnamefont {Park}}, \ and\ \bibinfo {author}
  {\bibfnamefont {S.-W.}\ \bibnamefont {Cheong}},\ }\href {\doibase
  10.1103/PhysRevB.80.100402} {\bibfield  {journal} {\bibinfo  {journal} {Phys.
  Rev. B}\ }\textbf {\bibinfo {volume} {80}},\ \bibinfo {pages} {100402}
  (\bibinfo {year} {2009})}\BibitemShut {NoStop}%
\bibitem [{\citenamefont {Furukawa}\ \emph {et~al.}(2010)\citenamefont
  {Furukawa}, \citenamefont {Sato},\ and\ \citenamefont
  {Onoda}}]{PhysRevLett.105.257205}%
  \BibitemOpen
  \bibfield  {author} {\bibinfo {author} {\bibfnamefont {S.}~\bibnamefont
  {Furukawa}}, \bibinfo {author} {\bibfnamefont {M.}~\bibnamefont {Sato}}, \
  and\ \bibinfo {author} {\bibfnamefont {S.}~\bibnamefont {Onoda}},\ }\href
  {\doibase 10.1103/PhysRevLett.105.257205} {\bibfield  {journal} {\bibinfo
  {journal} {Phys. Rev. Lett.}\ }\textbf {\bibinfo {volume} {105}},\ \bibinfo
  {pages} {257205} (\bibinfo {year} {2010})}\BibitemShut {NoStop}%
\bibitem [{\citenamefont {Gunnar}(2002)}]{Gunnar:2002aa}%
  \BibitemOpen
  \bibfield  {author} {\bibinfo {author} {\bibfnamefont {J.}~\bibnamefont
  {Gunnar}},\ }\href {\doibase
  doi:10.1002/1521-3927(20020301)23:4<227::AID-MARC227>3.0.CO;2-D} {\bibfield
  {journal} {\bibinfo  {journal} {Macromolecular Rapid Communications}\
  }\textbf {\bibinfo {volume} {23}},\ \bibinfo {pages} {227} (\bibinfo {year}
  {2002})}\BibitemShut {NoStop}%
\bibitem [{\citenamefont {Weil}\ and\ \citenamefont
  {Bolton}(2007)}]{weil2007electron}%
  \BibitemOpen
  \bibfield  {author} {\bibinfo {author} {\bibfnamefont {J.}~\bibnamefont
  {Weil}}\ and\ \bibinfo {author} {\bibfnamefont {J.}~\bibnamefont {Bolton}},\
  }\href {https://books.google.co.jp/books?id=qjLpMw9ZgPIC} {\emph {\bibinfo
  {title} {{Electron Paramagnetic Resonance: Elementary Theory and Practical
  Applications}}}}\ (\bibinfo  {publisher} {Wiley},\ \bibinfo {year}
  {2007})\BibitemShut {NoStop}%
\bibitem [{\citenamefont {Halliwell}\ and\ \citenamefont
  {Gutteridge}(2015)}]{halliwell2015free}%
  \BibitemOpen
  \bibfield  {author} {\bibinfo {author} {\bibfnamefont {B.}~\bibnamefont
  {Halliwell}}\ and\ \bibinfo {author} {\bibfnamefont {J.}~\bibnamefont
  {Gutteridge}},\ }\href {https://books.google.co.jp/books?id=3DlKCgAAQBAJ}
  {\emph {\bibinfo {title} {{Free Radicals in Biology and Medicine}}}}\
  (\bibinfo  {publisher} {Oxford University Press},\ \bibinfo {year}
  {2015})\BibitemShut {NoStop}%
\bibitem [{\citenamefont {Nagy}\ \emph {et~al.}(2011)\citenamefont {Nagy},
  \citenamefont {Quintavalle}, \citenamefont {Feh{\'e}r},\ and\ \citenamefont
  {J{\'a}nossy}}]{Nagy:2011aa}%
  \BibitemOpen
  \bibfield  {author} {\bibinfo {author} {\bibfnamefont {K.~L.}\ \bibnamefont
  {Nagy}}, \bibinfo {author} {\bibfnamefont {D.}~\bibnamefont {Quintavalle}},
  \bibinfo {author} {\bibfnamefont {T.}~\bibnamefont {Feh{\'e}r}}, \ and\
  \bibinfo {author} {\bibfnamefont {A.}~\bibnamefont {J{\'a}nossy}},\ }\href
  {\doibase 10.1007/s00723-010-0182-4} {\bibfield  {journal} {\bibinfo
  {journal} {Applied Magnetic Resonance}\ }\textbf {\bibinfo {volume} {40}},\
  \bibinfo {pages} {47} (\bibinfo {year} {2011})}\BibitemShut {NoStop}%
\bibitem [{\citenamefont {Chipman}(2005)}]{water}%
  \BibitemOpen
  \bibfield  {author} {\bibinfo {author} {\bibfnamefont {D.~M.}\ \bibnamefont
  {Chipman}},\ }\href@noop {} {\bibfield  {journal} {\bibinfo  {journal} {The
  Journal of Chemical Physics}\ }\textbf {\bibinfo {volume} {122}} (\bibinfo
  {year} {2005})}\BibitemShut {NoStop}%
\bibitem [{\citenamefont {Barra}\ \emph {et~al.}(1997)\citenamefont {Barra},
  \citenamefont {Gatteschi},\ and\ \citenamefont {Sessoli}}]{PhysRevB.56.8192}%
  \BibitemOpen
  \bibfield  {author} {\bibinfo {author} {\bibfnamefont {A.~L.}\ \bibnamefont
  {Barra}}, \bibinfo {author} {\bibfnamefont {D.}~\bibnamefont {Gatteschi}}, \
  and\ \bibinfo {author} {\bibfnamefont {R.}~\bibnamefont {Sessoli}},\ }\href
  {\doibase 10.1103/PhysRevB.56.8192} {\bibfield  {journal} {\bibinfo
  {journal} {Phys. Rev. B}\ }\textbf {\bibinfo {volume} {56}},\ \bibinfo
  {pages} {8192} (\bibinfo {year} {1997})}\BibitemShut {NoStop}%
\bibitem [{\citenamefont {Edwards}\ \emph {et~al.}(2003)\citenamefont
  {Edwards}, \citenamefont {Maccagnano}, \citenamefont {Yang}, \citenamefont
  {Hill}, \citenamefont {Wernsdorfer}, \citenamefont {Hendrickson},\ and\
  \citenamefont {Christou}}]{Edwards:2003aa}%
  \BibitemOpen
  \bibfield  {author} {\bibinfo {author} {\bibfnamefont {R.~S.}\ \bibnamefont
  {Edwards}}, \bibinfo {author} {\bibfnamefont {S.}~\bibnamefont {Maccagnano}},
  \bibinfo {author} {\bibfnamefont {E.~C.}\ \bibnamefont {Yang}}, \bibinfo
  {author} {\bibfnamefont {S.}~\bibnamefont {Hill}}, \bibinfo {author}
  {\bibfnamefont {W.}~\bibnamefont {Wernsdorfer}}, \bibinfo {author}
  {\bibfnamefont {D.}~\bibnamefont {Hendrickson}}, \ and\ \bibinfo {author}
  {\bibfnamefont {G.}~\bibnamefont {Christou}},\ }\bibfield  {booktitle} {\emph
  {\bibinfo {booktitle} {Journal of Applied Physics}},\ }\href {\doibase
  10.1063/1.1540050} {\bibfield  {journal} {\bibinfo  {journal} {Journal of
  Applied Physics}\ }\textbf {\bibinfo {volume} {93}},\ \bibinfo {pages} {7807}
  (\bibinfo {year} {2003})}\BibitemShut {NoStop}%
\bibitem [{\citenamefont {Yang}\ \emph {et~al.}(2005)\citenamefont {Yang},
  \citenamefont {Kirman}, \citenamefont {Lawrence}, \citenamefont {Zakharov},
  \citenamefont {Rheingold}, \citenamefont {Hill},\ and\ \citenamefont
  {Hendrickson}}]{Yang:2005aa}%
  \BibitemOpen
  \bibfield  {author} {\bibinfo {author} {\bibfnamefont {E.-C.}\ \bibnamefont
  {Yang}}, \bibinfo {author} {\bibfnamefont {C.}~\bibnamefont {Kirman}},
  \bibinfo {author} {\bibfnamefont {J.}~\bibnamefont {Lawrence}}, \bibinfo
  {author} {\bibfnamefont {L.~N.}\ \bibnamefont {Zakharov}}, \bibinfo {author}
  {\bibfnamefont {A.~L.}\ \bibnamefont {Rheingold}}, \bibinfo {author}
  {\bibfnamefont {S.}~\bibnamefont {Hill}}, \ and\ \bibinfo {author}
  {\bibfnamefont {D.~N.}\ \bibnamefont {Hendrickson}},\ }\bibfield  {booktitle}
  {\emph {\bibinfo {booktitle} {{Inorganic Chemistry}}},\ }\href {\doibase
  10.1021/ic0482279} {\bibfield  {journal} {\bibinfo  {journal} {Inorganic
  Chemistry}\ }\textbf {\bibinfo {volume} {44}},\ \bibinfo {pages} {3827}
  (\bibinfo {year} {2005})}\BibitemShut {NoStop}%
\bibitem [{\citenamefont {Ruamps}\ \emph {et~al.}(2013)\citenamefont {Ruamps},
  \citenamefont {Maurice}, \citenamefont {Batchelor}, \citenamefont
  {Boggio-Pasqua}, \citenamefont {Guillot}, \citenamefont {Barra},
  \citenamefont {Liu}, \citenamefont {Bendeif}, \citenamefont {Pillet},
  \citenamefont {Hill}, \citenamefont {Mallah},\ and\ \citenamefont
  {Guih{\'e}ry}}]{doi:10.1021/ja308146e}%
  \BibitemOpen
  \bibfield  {author} {\bibinfo {author} {\bibfnamefont {R.}~\bibnamefont
  {Ruamps}}, \bibinfo {author} {\bibfnamefont {R.}~\bibnamefont {Maurice}},
  \bibinfo {author} {\bibfnamefont {L.}~\bibnamefont {Batchelor}}, \bibinfo
  {author} {\bibfnamefont {M.}~\bibnamefont {Boggio-Pasqua}}, \bibinfo {author}
  {\bibfnamefont {R.}~\bibnamefont {Guillot}}, \bibinfo {author} {\bibfnamefont
  {A.~L.}\ \bibnamefont {Barra}}, \bibinfo {author} {\bibfnamefont
  {J.}~\bibnamefont {Liu}}, \bibinfo {author} {\bibfnamefont {E.-E.}\
  \bibnamefont {Bendeif}}, \bibinfo {author} {\bibfnamefont {S.}~\bibnamefont
  {Pillet}}, \bibinfo {author} {\bibfnamefont {S.}~\bibnamefont {Hill}},
  \bibinfo {author} {\bibfnamefont {T.}~\bibnamefont {Mallah}}, \ and\ \bibinfo
  {author} {\bibfnamefont {N.}~\bibnamefont {Guih{\'e}ry}},\ }\href {\doibase
  10.1021/ja308146e} {\bibfield  {journal} {\bibinfo  {journal} {Journal of the
  American Chemical Society}\ }\textbf {\bibinfo {volume} {135}},\ \bibinfo
  {pages} {3017} (\bibinfo {year} {2013})},\ \bibinfo {note} {pMID:
  23346898}\BibitemShut {NoStop}%
\bibitem [{\citenamefont {Marriott}\ \emph {et~al.}(2015)\citenamefont
  {Marriott}, \citenamefont {Bhaskaran}, \citenamefont {Wilson}, \citenamefont
  {Medarde}, \citenamefont {Ochsenbein}, \citenamefont {Hill},\ and\
  \citenamefont {Murrie}}]{C5SC02854J}%
  \BibitemOpen
  \bibfield  {author} {\bibinfo {author} {\bibfnamefont {K.~E.~R.}\
  \bibnamefont {Marriott}}, \bibinfo {author} {\bibfnamefont {L.}~\bibnamefont
  {Bhaskaran}}, \bibinfo {author} {\bibfnamefont {C.}~\bibnamefont {Wilson}},
  \bibinfo {author} {\bibfnamefont {M.}~\bibnamefont {Medarde}}, \bibinfo
  {author} {\bibfnamefont {S.~T.}\ \bibnamefont {Ochsenbein}}, \bibinfo
  {author} {\bibfnamefont {S.}~\bibnamefont {Hill}}, \ and\ \bibinfo {author}
  {\bibfnamefont {M.}~\bibnamefont {Murrie}},\ }\href {\doibase
  10.1039/C5SC02854J} {\bibfield  {journal} {\bibinfo  {journal} {Chem. Sci.}\
  }\textbf {\bibinfo {volume} {6}},\ \bibinfo {pages} {6823} (\bibinfo {year}
  {2015})}\BibitemShut {NoStop}%
\bibitem [{\citenamefont {Roth}\ \emph {et~al.}(2009)\citenamefont {Roth},
  \citenamefont {Br{\"u}ne}, \citenamefont {Buhmann}, \citenamefont
  {Molenkamp}, \citenamefont {Maciejko}, \citenamefont {Qi},\ and\
  \citenamefont {Zhang}}]{Roth294}%
  \BibitemOpen
  \bibfield  {author} {\bibinfo {author} {\bibfnamefont {A.}~\bibnamefont
  {Roth}}, \bibinfo {author} {\bibfnamefont {C.}~\bibnamefont {Br{\"u}ne}},
  \bibinfo {author} {\bibfnamefont {H.}~\bibnamefont {Buhmann}}, \bibinfo
  {author} {\bibfnamefont {L.~W.}\ \bibnamefont {Molenkamp}}, \bibinfo {author}
  {\bibfnamefont {J.}~\bibnamefont {Maciejko}}, \bibinfo {author}
  {\bibfnamefont {X.-L.}\ \bibnamefont {Qi}}, \ and\ \bibinfo {author}
  {\bibfnamefont {S.-C.}\ \bibnamefont {Zhang}},\ }\href {\doibase
  10.1126/science.1174736} {\bibfield  {journal} {\bibinfo  {journal}
  {Science}\ }\textbf {\bibinfo {volume} {325}},\ \bibinfo {pages} {294}
  (\bibinfo {year} {2009})}\BibitemShut {NoStop}%
\bibitem [{\citenamefont {Chen}\ \emph {et~al.}(2014)\citenamefont {Chen},
  \citenamefont {Jiang}, \citenamefont {Chen}, \citenamefont {Zhu},
  \citenamefont {Zhou}, \citenamefont {Dong},\ and\ \citenamefont
  {Chan}}]{Chen:2014aa}%
  \BibitemOpen
  \bibfield  {author} {\bibinfo {author} {\bibfnamefont {W.-J.}\ \bibnamefont
  {Chen}}, \bibinfo {author} {\bibfnamefont {S.-J.}\ \bibnamefont {Jiang}},
  \bibinfo {author} {\bibfnamefont {X.-D.}\ \bibnamefont {Chen}}, \bibinfo
  {author} {\bibfnamefont {B.}~\bibnamefont {Zhu}}, \bibinfo {author}
  {\bibfnamefont {L.}~\bibnamefont {Zhou}}, \bibinfo {author} {\bibfnamefont
  {J.-W.}\ \bibnamefont {Dong}}, \ and\ \bibinfo {author} {\bibfnamefont
  {C.~T.}\ \bibnamefont {Chan}},\ }\href {http://dx.doi.org/10.1038/ncomms6782}
  {\bibfield  {journal} {\bibinfo  {journal} {Nature Communications}\ }\textbf
  {\bibinfo {volume} {5}},\ \bibinfo {pages} {5782 EP } (\bibinfo {year}
  {2014})}\BibitemShut {NoStop}%
\bibitem [{\citenamefont {S{\"u}sstrunk}\ and\ \citenamefont
  {Huber}(2015)}]{Susstrunk:2015aa}%
  \BibitemOpen
  \bibfield  {author} {\bibinfo {author} {\bibfnamefont {R.}~\bibnamefont
  {S{\"u}sstrunk}}\ and\ \bibinfo {author} {\bibfnamefont {S.~D.}\ \bibnamefont
  {Huber}},\ }\href
  {http://science.sciencemag.org/content/349/6243/47.abstract} {\bibfield
  {journal} {\bibinfo  {journal} {Science}\ }\textbf {\bibinfo {volume}
  {349}},\ \bibinfo {pages} {47} (\bibinfo {year} {2015})}\BibitemShut
  {NoStop}%
\bibitem [{\citenamefont {Ito}\ \emph {et~al.}(2011)\citenamefont {Ito},
  \citenamefont {Furuya}, \citenamefont {Shibata}, \citenamefont {Kashiwaya},
  \citenamefont {Yamaguchi}, \citenamefont {Akazaki}, \citenamefont {Tamura},
  \citenamefont {Ootuka},\ and\ \citenamefont
  {Nomura}}]{PhysRevLett.107.256803}%
  \BibitemOpen
  \bibfield  {author} {\bibinfo {author} {\bibfnamefont {H.}~\bibnamefont
  {Ito}}, \bibinfo {author} {\bibfnamefont {K.}~\bibnamefont {Furuya}},
  \bibinfo {author} {\bibfnamefont {Y.}~\bibnamefont {Shibata}}, \bibinfo
  {author} {\bibfnamefont {S.}~\bibnamefont {Kashiwaya}}, \bibinfo {author}
  {\bibfnamefont {M.}~\bibnamefont {Yamaguchi}}, \bibinfo {author}
  {\bibfnamefont {T.}~\bibnamefont {Akazaki}}, \bibinfo {author} {\bibfnamefont
  {H.}~\bibnamefont {Tamura}}, \bibinfo {author} {\bibfnamefont
  {Y.}~\bibnamefont {Ootuka}}, \ and\ \bibinfo {author} {\bibfnamefont
  {S.}~\bibnamefont {Nomura}},\ }\href {\doibase
  10.1103/PhysRevLett.107.256803} {\bibfield  {journal} {\bibinfo  {journal}
  {Phys. Rev. Lett.}\ }\textbf {\bibinfo {volume} {107}},\ \bibinfo {pages}
  {256803} (\bibinfo {year} {2011})}\BibitemShut {NoStop}%
\bibitem [{\citenamefont {Lai}\ \emph {et~al.}(2011)\citenamefont {Lai},
  \citenamefont {Kundhikanjana}, \citenamefont {Kelly}, \citenamefont {Shen},
  \citenamefont {Shabani},\ and\ \citenamefont
  {Shayegan}}]{PhysRevLett.107.176809}%
  \BibitemOpen
  \bibfield  {author} {\bibinfo {author} {\bibfnamefont {K.}~\bibnamefont
  {Lai}}, \bibinfo {author} {\bibfnamefont {W.}~\bibnamefont {Kundhikanjana}},
  \bibinfo {author} {\bibfnamefont {M.~A.}\ \bibnamefont {Kelly}}, \bibinfo
  {author} {\bibfnamefont {Z.-X.}\ \bibnamefont {Shen}}, \bibinfo {author}
  {\bibfnamefont {J.}~\bibnamefont {Shabani}}, \ and\ \bibinfo {author}
  {\bibfnamefont {M.}~\bibnamefont {Shayegan}},\ }\href {\doibase
  10.1103/PhysRevLett.107.176809} {\bibfield  {journal} {\bibinfo  {journal}
  {Phys. Rev. Lett.}\ }\textbf {\bibinfo {volume} {107}},\ \bibinfo {pages}
  {176809} (\bibinfo {year} {2011})}\BibitemShut {NoStop}%
\bibitem [{\citenamefont {Nowack}\ \emph {et~al.}(2013)\citenamefont {Nowack},
  \citenamefont {Spanton}, \citenamefont {Baenninger}, \citenamefont
  {K{\"o}nig}, \citenamefont {Kirtley}, \citenamefont {Kalisky}, \citenamefont
  {Ames}, \citenamefont {Leubner}, \citenamefont {Br{\"u}ne}, \citenamefont
  {Buhmann}, \citenamefont {Molenkamp}, \citenamefont {Goldhaber-Gordon},\ and\
  \citenamefont {Moler}}]{Nowack:2013aa}%
  \BibitemOpen
  \bibfield  {author} {\bibinfo {author} {\bibfnamefont {K.~C.}\ \bibnamefont
  {Nowack}}, \bibinfo {author} {\bibfnamefont {E.~M.}\ \bibnamefont {Spanton}},
  \bibinfo {author} {\bibfnamefont {M.}~\bibnamefont {Baenninger}}, \bibinfo
  {author} {\bibfnamefont {M.}~\bibnamefont {K{\"o}nig}}, \bibinfo {author}
  {\bibfnamefont {J.~R.}\ \bibnamefont {Kirtley}}, \bibinfo {author}
  {\bibfnamefont {B.}~\bibnamefont {Kalisky}}, \bibinfo {author} {\bibfnamefont
  {C.}~\bibnamefont {Ames}}, \bibinfo {author} {\bibfnamefont {P.}~\bibnamefont
  {Leubner}}, \bibinfo {author} {\bibfnamefont {C.}~\bibnamefont {Br{\"u}ne}},
  \bibinfo {author} {\bibfnamefont {H.}~\bibnamefont {Buhmann}}, \bibinfo
  {author} {\bibfnamefont {L.~W.}\ \bibnamefont {Molenkamp}}, \bibinfo {author}
  {\bibfnamefont {D.}~\bibnamefont {Goldhaber-Gordon}}, \ and\ \bibinfo
  {author} {\bibfnamefont {K.~A.}\ \bibnamefont {Moler}},\ }\href
  {http://dx.doi.org/10.1038/nmat3682} {\bibfield  {journal} {\bibinfo
  {journal} {Nature Materials}\ }\textbf {\bibinfo {volume} {12}},\ \bibinfo
  {pages} {787 EP } (\bibinfo {year} {2013})}\BibitemShut {NoStop}%
\bibitem [{\citenamefont {Resta}(2010)}]{0953-8984-22-12-123201}%
  \BibitemOpen
  \bibfield  {author} {\bibinfo {author} {\bibfnamefont {R.}~\bibnamefont
  {Resta}},\ }\href {http://stacks.iop.org/0953-8984/22/i=12/a=123201}
  {\bibfield  {journal} {\bibinfo  {journal} {Journal of Physics: Condensed
  Matter}\ }\textbf {\bibinfo {volume} {22}},\ \bibinfo {pages} {123201}
  (\bibinfo {year} {2010})}\BibitemShut {NoStop}%
\bibitem [{\citenamefont {Thonhauser}(2011)}]{doi:10.1142/S0217979211058912}%
  \BibitemOpen
  \bibfield  {author} {\bibinfo {author} {\bibfnamefont {T.}~\bibnamefont
  {Thonhauser}},\ }\href {\doibase 10.1142/S0217979211058912} {\bibfield
  {journal} {\bibinfo  {journal} {International Journal of Modern Physics B}\
  }\textbf {\bibinfo {volume} {25}},\ \bibinfo {pages} {1429} (\bibinfo {year}
  {2011})}\BibitemShut {NoStop}%
\bibitem [{Note6()}]{Note6}%
  \BibitemOpen
  \bibinfo {note} {More precisely, the orbital magnetization is a sum of the
  surface and bulk contributions. The latter which we ignored comes from the
  microscopic circulating motions of electrons around each atom. Since these
  motions do not couple to the azimuthal electric field, the bulk contribution
  to the orbital magnetization would be negligible in the present
  case.}\BibitemShut {Stop}%
\bibitem [{\citenamefont {Fausti}\ \emph {et~al.}(2011)\citenamefont {Fausti},
  \citenamefont {Tobey}, \citenamefont {Dean}, \citenamefont {Kaiser},
  \citenamefont {Dienst}, \citenamefont {Hoffmann}, \citenamefont {Pyon},
  \citenamefont {Takayama}, \citenamefont {Takagi},\ and\ \citenamefont
  {Cavalleri}}]{Fausti189}%
  \BibitemOpen
  \bibfield  {author} {\bibinfo {author} {\bibfnamefont {D.}~\bibnamefont
  {Fausti}}, \bibinfo {author} {\bibfnamefont {R.~I.}\ \bibnamefont {Tobey}},
  \bibinfo {author} {\bibfnamefont {N.}~\bibnamefont {Dean}}, \bibinfo {author}
  {\bibfnamefont {S.}~\bibnamefont {Kaiser}}, \bibinfo {author} {\bibfnamefont
  {A.}~\bibnamefont {Dienst}}, \bibinfo {author} {\bibfnamefont {M.~C.}\
  \bibnamefont {Hoffmann}}, \bibinfo {author} {\bibfnamefont {S.}~\bibnamefont
  {Pyon}}, \bibinfo {author} {\bibfnamefont {T.}~\bibnamefont {Takayama}},
  \bibinfo {author} {\bibfnamefont {H.}~\bibnamefont {Takagi}}, \ and\ \bibinfo
  {author} {\bibfnamefont {A.}~\bibnamefont {Cavalleri}},\ }\href {\doibase
  10.1126/science.1197294} {\bibfield  {journal} {\bibinfo  {journal}
  {Science}\ }\textbf {\bibinfo {volume} {331}},\ \bibinfo {pages} {189}
  (\bibinfo {year} {2011})}\BibitemShut {NoStop}%
\bibitem [{\citenamefont
  {Cavalleri}(2018)}]{doi:10.1080/00107514.2017.1406623}%
  \BibitemOpen
  \bibfield  {author} {\bibinfo {author} {\bibfnamefont {A.}~\bibnamefont
  {Cavalleri}},\ }\href {\doibase 10.1080/00107514.2017.1406623} {\bibfield
  {journal} {\bibinfo  {journal} {Contemporary Physics}\ }\textbf {\bibinfo
  {volume} {59}},\ \bibinfo {pages} {31} (\bibinfo {year} {2018})},\ \Eprint
  {http://arxiv.org/abs/https://doi.org/10.1080/00107514.2017.1406623}
  {https://doi.org/10.1080/00107514.2017.1406623} \BibitemShut {NoStop}%
\bibitem [{\citenamefont {Takayoshi}\ \emph
  {et~al.}(2014{\natexlab{a}})\citenamefont {Takayoshi}, \citenamefont {Aoki},\
  and\ \citenamefont {Oka}}]{PhysRevB.90.085150}%
  \BibitemOpen
  \bibfield  {author} {\bibinfo {author} {\bibfnamefont {S.}~\bibnamefont
  {Takayoshi}}, \bibinfo {author} {\bibfnamefont {H.}~\bibnamefont {Aoki}}, \
  and\ \bibinfo {author} {\bibfnamefont {T.}~\bibnamefont {Oka}},\ }\href
  {\doibase 10.1103/PhysRevB.90.085150} {\bibfield  {journal} {\bibinfo
  {journal} {Phys. Rev. B}\ }\textbf {\bibinfo {volume} {90}},\ \bibinfo
  {pages} {085150} (\bibinfo {year} {2014}{\natexlab{a}})}\BibitemShut
  {NoStop}%
\bibitem [{\citenamefont {Takayoshi}\ \emph
  {et~al.}(2014{\natexlab{b}})\citenamefont {Takayoshi}, \citenamefont {Sato},\
  and\ \citenamefont {Oka}}]{PhysRevB.90.214413}%
  \BibitemOpen
  \bibfield  {author} {\bibinfo {author} {\bibfnamefont {S.}~\bibnamefont
  {Takayoshi}}, \bibinfo {author} {\bibfnamefont {M.}~\bibnamefont {Sato}}, \
  and\ \bibinfo {author} {\bibfnamefont {T.}~\bibnamefont {Oka}},\ }\href
  {\doibase 10.1103/PhysRevB.90.214413} {\bibfield  {journal} {\bibinfo
  {journal} {Phys. Rev. B}\ }\textbf {\bibinfo {volume} {90}},\ \bibinfo
  {pages} {214413} (\bibinfo {year} {2014}{\natexlab{b}})}\BibitemShut
  {NoStop}%
\bibitem [{\citenamefont {Goldman}\ and\ \citenamefont
  {Dalibard}(2014)}]{PhysRevX.4.031027}%
  \BibitemOpen
  \bibfield  {author} {\bibinfo {author} {\bibfnamefont {N.}~\bibnamefont
  {Goldman}}\ and\ \bibinfo {author} {\bibfnamefont {J.}~\bibnamefont
  {Dalibard}},\ }\href {\doibase 10.1103/PhysRevX.4.031027} {\bibfield
  {journal} {\bibinfo  {journal} {Phys. Rev. X}\ }\textbf {\bibinfo {volume}
  {4}},\ \bibinfo {pages} {031027} (\bibinfo {year} {2014})}\BibitemShut
  {NoStop}%
\bibitem [{\citenamefont {Oka}\ and\ \citenamefont
  {Aoki}(2009)}]{PhysRevB.79.081406}%
  \BibitemOpen
  \bibfield  {author} {\bibinfo {author} {\bibfnamefont {T.}~\bibnamefont
  {Oka}}\ and\ \bibinfo {author} {\bibfnamefont {H.}~\bibnamefont {Aoki}},\
  }\href {\doibase 10.1103/PhysRevB.79.081406} {\bibfield  {journal} {\bibinfo
  {journal} {Phys. Rev. B}\ }\textbf {\bibinfo {volume} {79}},\ \bibinfo
  {pages} {081406} (\bibinfo {year} {2009})}\BibitemShut {NoStop}%
\bibitem [{\citenamefont {Wang}\ \emph {et~al.}(2013)\citenamefont {Wang},
  \citenamefont {Steinberg}, \citenamefont {Jarillo-Herrero},\ and\
  \citenamefont {Gedik}}]{Wang453}%
  \BibitemOpen
  \bibfield  {author} {\bibinfo {author} {\bibfnamefont {Y.~H.}\ \bibnamefont
  {Wang}}, \bibinfo {author} {\bibfnamefont {H.}~\bibnamefont {Steinberg}},
  \bibinfo {author} {\bibfnamefont {P.}~\bibnamefont {Jarillo-Herrero}}, \ and\
  \bibinfo {author} {\bibfnamefont {N.}~\bibnamefont {Gedik}},\ }\href
  {\doibase 10.1126/science.1239834} {\bibfield  {journal} {\bibinfo  {journal}
  {Science}\ }\textbf {\bibinfo {volume} {342}},\ \bibinfo {pages} {453}
  (\bibinfo {year} {2013})}\BibitemShut {NoStop}%
\bibitem [{\citenamefont {McIver}\ \emph {et~al.}(2018)\citenamefont {McIver},
  \citenamefont {Schulte}, \citenamefont {Stein}, \citenamefont {Matsuyama},
  \citenamefont {Jotzu}, \citenamefont {Meier},\ and\ \citenamefont
  {Cavalleri}}]{arXiv:1811.03522}%
  \BibitemOpen
  \bibfield  {author} {\bibinfo {author} {\bibfnamefont {J.}~\bibnamefont
  {McIver}}, \bibinfo {author} {\bibfnamefont {B.}~\bibnamefont {Schulte}},
  \bibinfo {author} {\bibfnamefont {F.-U.}\ \bibnamefont {Stein}}, \bibinfo
  {author} {\bibfnamefont {T.}~\bibnamefont {Matsuyama}}, \bibinfo {author}
  {\bibfnamefont {G.}~\bibnamefont {Jotzu}}, \bibinfo {author} {\bibfnamefont
  {G.}~\bibnamefont {Meier}}, \ and\ \bibinfo {author} {\bibfnamefont
  {A.}~\bibnamefont {Cavalleri}},\ }\href@noop {} {\bibfield  {journal}
  {\bibinfo  {journal} {arXiv:1811.03522}\ } (\bibinfo {year}
  {2018})}\BibitemShut {NoStop}%
\bibitem [{\citenamefont {Dzyaloshinsky}(1958)}]{DZYALOSHINSKY1958241}%
  \BibitemOpen
  \bibfield  {author} {\bibinfo {author} {\bibfnamefont {I.}~\bibnamefont
  {Dzyaloshinsky}},\ }\href {\doibase
  http://dx.doi.org/10.1016/0022-3697(58)90076-3} {\bibfield  {journal}
  {\bibinfo  {journal} {J. Phys. Chem. Solids}\ }\textbf {\bibinfo {volume}
  {4}},\ \bibinfo {pages} {241 } (\bibinfo {year} {1958})}\BibitemShut
  {NoStop}%
\bibitem [{\citenamefont {Moriya}(1960)}]{PhysRev.120.91}%
  \BibitemOpen
  \bibfield  {author} {\bibinfo {author} {\bibfnamefont {T.}~\bibnamefont
  {Moriya}},\ }\href {\doibase 10.1103/PhysRev.120.91} {\bibfield  {journal}
  {\bibinfo  {journal} {Phys. Rev.}\ }\textbf {\bibinfo {volume} {120}},\
  \bibinfo {pages} {91} (\bibinfo {year} {1960})}\BibitemShut {NoStop}%
\bibitem [{\citenamefont {Maekawa}\ \emph {et~al.}(2012)\citenamefont
  {Maekawa}, \citenamefont {Valenzuela}, \citenamefont {Saitoh},\ and\
  \citenamefont {Kimura}}]{spincurrentbook}%
  \BibitemOpen
  \bibinfo {editor} {\bibfnamefont {S.}~\bibnamefont {Maekawa}}, \bibinfo
  {editor} {\bibfnamefont {S.~O.}\ \bibnamefont {Valenzuela}}, \bibinfo
  {editor} {\bibfnamefont {E.}~\bibnamefont {Saitoh}}, \ and\ \bibinfo {editor}
  {\bibfnamefont {T.}~\bibnamefont {Kimura}},\ eds.,\ \href@noop {} {\emph
  {\bibinfo {title} {{Spin Current}}}}\ (\bibinfo  {publisher} {Oxford
  University Press},\ \bibinfo {year} {2012})\BibitemShut {NoStop}%
\end{thebibliography}%

\appendix

\section{ Bloch equation for magnetic resonance}\label{app: bloch}
In this appendix, we review the Bloch equation and derive Eq.~\eqref{chi} of the main text. The Bloch equation is an equation of motion of a magnetization vector $\bm{M}(t) = (M^x(t), M^y(t), M^z(t))$ in the presence of an external magnetic field and relaxations. In the initial state, a static external field: $\bm{B} = (0, 0, H_0)$ is applied and the magnetization is along that: $\bm{M}(0) = (0, 0, M_0)$. At $t = 0$, an oscillating magnetic field is applied on top of the static field: $\bm{B}(t)=(B^x(t), B^y(t), H_0)$. Then the following Bloch equation describes the time evolution of the magnetization as
\begin{align} \label{Bloch}
\frac{M^x(t)}{dt} &= \gamma [\bm{M}(t)\times\bm{B}(t)]^x - \frac{M^x(t)}{T_2}\nonumber \\
\frac{M^y(t)}{dt} &= \gamma [\bm{M}(t)\times\bm{B}(t)]^y - \frac{M^y(t)}{T_2}\nonumber \\
\frac{M^z(t)}{dt} &= \gamma [\bm{M}(t)\times\bm{B}(t)]^z - \frac{M^z(t) - M_0}{T_1}.
\end{align}
Here $\gamma$ is the gyromagnetic ratio and $T_1$, $T_2$ are longitudinal and transverse relaxation times respectively. 

Let us consider a linearly polarized field: $\bm{B}(t) = B \cos(\omega t) \hat{x} + H_0 \hat{z}$. The oscillating component is decomposed into the left-handed and right-handed ones: $\Delta\bm{B}^l(t) = B_l( \cos(\omega t), -i \sin(\omega t), 0)$ and $\Delta\bm{B}^r(t) = B_r( \cos(\omega t), i \sin(\omega t), 0)$. Near the magnetic resonance, the left-handed component $\Delta\bm{B}^l(t)$ is relatively unimportant and we can take $\bm{B}(t) = \Delta\bm{B}^r(t) + H_0 \hat{z}$.

To obtain the steady-state solution, we move into a frame rotating around the $z$ axis with the angular velocity $\omega$. In the frame, the external field becomes static: $\bm{B}(t) \rightarrow B_r\hat{x} + (H_0 + \frac{\omega}{\gamma})\hat{z} = B_r \hat{x} + B_{\rm eff}^z \hat{z}$. If we write the magnetization in the frame as $\widetilde{\bm{M}}(t)$, the Bloch equation reads:
\begin{align} \label{Bloch-rot}
\frac{\widetilde{M}^x(t)}{dt} &=  \gamma \widetilde{M}^y(t) B^z_{\rm eff} - \frac{\widetilde{M}^x(t)}{T_2} \\
\frac{\widetilde{M}^y(t)}{dt} &=   \gamma \left[\widetilde{M}^z(t) B_r - \widetilde{M}^x(t) B^z_{\rm eff}\right] - \frac{\widetilde{M}^y(t)}{T_2} \\
\frac{\widetilde{M}^z(t)}{dt} &=  -\gamma \widetilde{M}^y(t) B_r  - \frac{\widetilde{M}^z(t) - M_0}{T_1},
\end{align}
and the steady-state solution is obtained by setting the left-hand sides to be zero. The $y$ component of the magnetization in the steady state is
\begin{align}
\widetilde{M}^y = \chi_0 \gamma H_0 T_2 \frac{1}{1 + (\gamma B_r)^2 T_1 T_2 + (\omega - \gamma H_0)^2 T_2^2}B_r,
\end{align}
where $\chi_0 = M_0/H_0$ is the static magnetic susceptibility. Assuming that the oscillating field is sufficiently weak, we drop the second term in the denominator:
\begin{align}
\widetilde{M}^y &\simeq \frac{\chi_0}{2} \gamma H_0 T_2 \frac{2 B_r}{1 + (\omega - \gamma H_0)^2 T_2^2}.
\end{align}
The imaginary part of the magnetic susceptibility is calculated to be $\widetilde{M}^y/B = \widetilde{M}^y/(2 B_r)$.

\section{Analytical result of the absorption by the edge state}\label{app: full}

Here we present the analytical expression of the integration \eqref{abs_edge}:

\begin{align}\label{fullform}
\alpha(w) &\propto \int_0^R \frac{1}{w} e^{2\frac{\rho-R}{\xi}}\left(\frac{\rho}{w}\right)^2e^{-2\rho^2/w^2}  \rho d\rho  \nonumber \\
&=\frac{e^{-2 R \left(\frac{1}{\xi }+\frac{R}{w^2}\right)}}{32 \xi ^3 w}
   (\sqrt{2 \pi } w^3 \left(3 \xi ^2+w^2\right) e^{\frac{2
   R^2}{w^2}+\frac{w^2}{2 \xi ^2}}
   \left[\text{erf}\left(\frac{w}{\sqrt{2} \xi
   }\right)-\text{erf}\left(\frac{w^2-2 \xi  R}{\sqrt{2} \xi 
   w}\right)\right]\nonumber\\ &+2 \xi  w^2 e^{\frac{2 R^2}{w^2}} \left(2 \xi
   ^2+w^2\right)-2 \xi  e^{\frac{2 R}{\xi }} \left(4 \xi ^2 R^2+2
   \xi  w^2 (\xi +R)+w^4\right)),
\end{align}
where $\text{erf}(x)$ is the error function. Assuming that the localization length $\xi$ is sufficiently smaller than the beam width $w$ and the sample radius $R$, we can expand Eq.~\eqref{fullform} in terms of $\xi$ to obtain a simplified expression:
\begin{align}
\alpha(w) \propto \frac{\xi  R^3 e^{-\frac{2 R^2}{w^2}}}{2 w^3}-\frac{\xi ^2 R^2
   e^{-\frac{2 R^2}{w^2}} \left(3 w^2-4 R^2\right)}{4 w^5},
\end{align}
up to the sub-leading order. As the prefactor omitted here does not depend on the beam width $w$, we can use this formula as a fitting function for the experimental data to obtain the profile of the localized current.

\section{Derivation of Eq.~\eqref{llg}}\label{eq10}
\begin{figure}[htbp]
\centering
\includegraphics[width = 80mm]{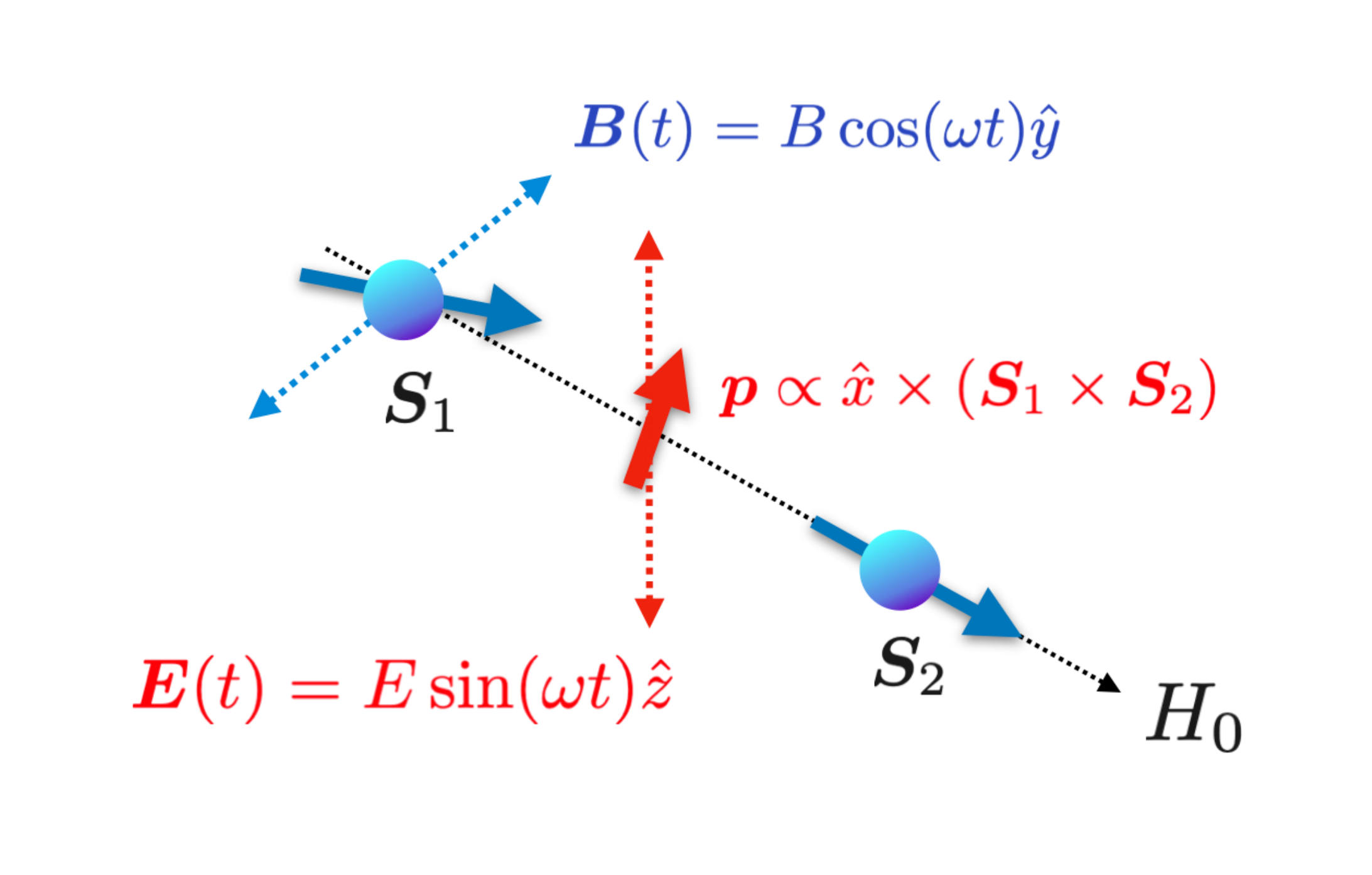}
\caption{Setup of the calculation for deriving Eq.~\eqref{llg} of the main text. The system is under a linearly polarized beam and a static magnetic field $H_0$ applied in the $x$ direction. The spin $\bm{S}_1$ couples with them through the Zeeman and magneto-electric couplings. For simplicity, we take $\bm{S}_2 = \hat{x}$. This assumption corresponds to ignoring the higher-order effects of the incident beam to the dynamics of the spin $\bm{S}_1$. The spin-induced electric polarization $\bm{p}\propto \hat{x}\times(\bm{S}_1\times\bm{S}_2)$ comes from the spin-current mechanism~\cite{PhysRevLett.95.057205}.}
       \label{mf_setup}
  \end{figure}

In this appendix, we derive Eq.~\eqref{llg}. Here we re-state our problem; a pair of (classical) spin coupled with a linearly polarized beam through the Zeeman and magneto-electric couplings. In Fig.~\ref{mf_setup}, we show the setup of our calculation.

The Hamiltonian we consider is the following one:
\begin{align}
    H &= - \bm{p}\cdot\bm{E}(t) -g\mu_B \bm{S}_1\cdot\bm{B}(t)- g \mu_B H_0 S_1^x \nonumber \\
    &= -\lambda_c  S_1^z E\sin(\omega t)  - g \mu_B S_1^y B \cos(\omega t) - g\mu_B H_0 S_1^x.
\end{align}
The dynamics of $\bm{S}_1$ is, in the framework of the Landau-Lifshitz equation~\cite{spincurrentbook}, determined by
\begin{align}\label{llg_formal}
    \frac{d\bm{S}_1}{d t} = -\gamma \bm{S}_1\times \bm{H}_{\rm eff},
\end{align}
where $\gamma = g\mu_B/\hbar$ is the gyromagnetic ratio and $\hbar \equiv 1$ is the Dirac's constant. Here the effective magnetic field $\bm{H}_{\rm eff} = - \frac{1}{\hbar \gamma} \frac{\partial H}{\partial \bm{S}_1 } $ is calculated to be
\begin{align}\label{eff_field}
    \bm{H}_{\rm eff} =  H_0 \hat{x} +   B\cos(\omega t) \hat{y} + \frac{\lambda_c E}{g\mu_B} \sin(\omega t)\hat{z}
\end{align}
Assuming that the oscillating fields is weak compared with $H_0$, below we use the approximation $S_1^x = 1$.

Under this assumption, Eq.~\eqref{llg_formal} is easily solved. The general solution of the dynamics of $S^y_1(t)$ is given as
\begin{align}
S_1^y(t) = \frac{\gamma^2 H_0 B + \omega \lambda_c E}{\gamma^2 H_0^2 - \omega^2}\cos(\omega t) + c_1 \cos(\sqrt{\gamma H_0}t) + c_2 \sin(\sqrt{\gamma H_0}t)
\end{align}
where $c_1$ and $c_2$ are constants. If we assume that the spin $\bm{S}_1$ was time-independent in the absence of both $E$ and $B$, we can set $c_1 = c_2 = 0$ and obtain Eq.~\eqref{llg} in the main text.

\section{Validation of the Floquet effective Hamiltonian}\label{floquet_numer}
Let us compare the effective description~\eqref{synth} and a numerical calculation. We take a pair of $s = 1/2$ quantum spins; $\bm{S}_1$ and $\bm{S}_2$. Since we are interested only in the validity of the Floquet effective Hamiltonian obtained with the Floquet-Magnus expansion, for simplicity, here we pin $\bm{S}_2$ to the $+z$ direction and ignore the time-independent terms (namely, we take $\hat{H}_0 = 0$). In this case, Eq.~\eqref{synth} is reduced to be 
\begin{align}\label{synth_0}
    H_F = -\frac{\sin\delta}{2 \omega}gg_{me}\mu_B E B S_1^y S_2^z  + O\left(\frac{1}{\omega^2}\right),
\end{align}
where $S_2^z = 1/2$. Here we apply the electric and magnetic fields in parallel. Namely, the optical axis of both the azimuthal and radial CVBs is the $z$ axis. The Hamiltonian~\eqref{synth_0} is equivalent to the Zeeman coupling with a synthetic magnetic field $\bm{H} = \frac{\sin\delta}{4 \omega}g_{me} E B S_2^z\hat{y}$. Since $\bm{S}_2$ is pinned to the $+z$ direction, this effective magnetic field favors a configuration with $\pi/2$ relative angle between the spins and works as a Dzyaloshinskii-Moriya interaction. 

To investigate the dynamical property of the driven spin $\bm{S}_1$, we take the initial condition $\bm{S}_1 = \hat{z}/2$. In this case, in the absence of any dissipation, after turning on the fields, the spin $\bm{S}_1$ perpetually precesses around the $\pm y$ direction for $\delta = \pm\pi/2$ ($\delta$ is the relative phase between the two beams as mentioned in the main text). We directly solve the Schr\"{o}dinger equation for the time-periodic Hamiltonian
\begin{align}
H(t) &=   -g\mu_B\bm{S}_1\cdot \bm{B}(t)- \bm{p}\cdot \bm{E}(t)\nonumber \\
 &= -(g\mu_B B) \cos(\omega t + \delta) S_1^z + (E\lambda_c) \cos(\omega t)S_1^x
\end{align}
and compare that with the expectation above based on the effective Floquet Hamiltonian. Here $\bm{p} = \lambda_c \hat{x}\times(\bm{S}_1\times\hat{z})$ is the electric polarization from the spin-current mechanism and $\bm{B}(t) = B\cos(\omega t + \delta)\hat{z}$ and $\bm{E}(t) = E\cos(\omega t)\hat{z}$ are electromagnetic fields from azimuthal and radial CVBs respectively.

\begin{figure}[htbp]
\centering
\includegraphics[width = 110mm]{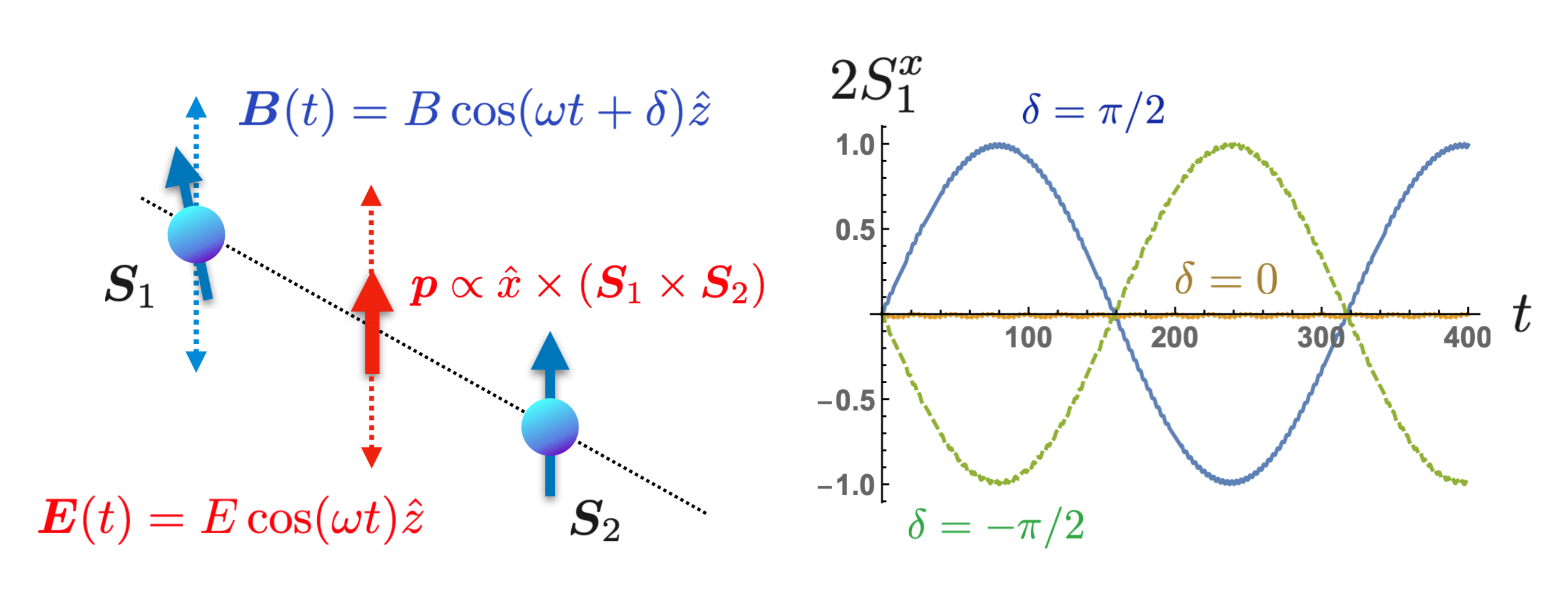}
\caption{Spin precession induced by the synthetic Dzyaloshinskii-Moriya interaction for three different values of the relative phase $\delta$. We take $\lambda_c E = g\mu_B B= 0.1$ and $\omega = 1$. The initial condition for $\bm{S}_1$ at $t = 0$ is $2\bm{S}_1 = \hat{z}$. The time is measured in the unit of $1/\omega$ }
       \label{CVB floquet}
  \end{figure}

The left panel of Fig.~\ref{CVB floquet} shows the setup of our calculation, and the right panel does the real-time evolution of $S_1^x$ for $\lambda_c E = g\mu_B B= 0.1$. We are measuring the energy in the unit of $\omega$. Since $\omega \gg \lambda_c E, g\mu_B B$, the high-frequency expansion should be a good approximation. The time-evolution shown in the right panel of Fig.~\ref{CVB floquet} is indeed consistent with the argument above. For $\delta =\pm \pi/2$, the spin precesses around $\pm y$ axis but for $\delta = 0$, we see no spin precession.

\end{document}